\documentclass[longauth]{aa}
\usepackage{url,hyperref}
\usepackage[varg]{txfonts}
\usepackage{graphicx}
\usepackage{footmisc}

\usepackage{amssymb} 
\usepackage{amsmath} 

\usepackage{xspace} 
\usepackage{natbib} 
\usepackage{caption}
\usepackage{subcaption}
\usepackage{color}    
\usepackage{placeins} 

\newcommand{\Fermi}{\textit{Fermi}\xspace}
\newcommand{\FermiLAT}{\textit{Fermi} LAT\xspace}

\newcommand{\He}{H.E.S.S.\xspace}

\newcommand{\HeII}{H.E.S.S.~II\xspace}

\newcommand{\Pk}{PKS\,2155$-$304\xspace}
\newcommand\simlt{\lower.5ex\hbox{$\; \buildrel < \over \sim
\;$}} \newcommand\simgt{\lower.5ex\hbox{$\; \buildrel > \over
\sim \;$}}

\newcommand{\degree}{\hbox{$^\circ$}\xspace}

\newcommand{\fixedbetahesssfvalues}{1.00^{+0.11}_{-0.15}}
\newcommand{\fixedbetahesssf}{\beta_\mathrm{SF}=\fixedbetahesssfvalues\xspace}
\newcommand{\fixedbetahesslsvalues}{1.10^{+0.14}_{-0.16}}
\newcommand{\fixedbetahessls}{\beta_\mathrm{LSP}=\fixedbetahesslsvalues\xspace}
\newcommand{\fixedbetahessmfvalues}{1.10^{+0.10}_{-0.13}}
\newcommand{\fixedbetahessmf}{\beta_\mathrm{MFVF}=\fixedbetahessmfvalues\xspace}
\newcommand{\fixedbetahessmeanvalue}{1.10^{+0.10}_{-0.13}}
\newcommand{\fixedbetahessmean}{\beta_\textrm{VHE}=\fixedbetahessmeanvalue\xspace}

\newcommand{\fixedbetafermisfvalues}{1.20^{+0.22}_{-0.31}}
\newcommand{\fixedbetafermisf}{\beta_\mathrm{SF}=\fixedbetafermisfvalues\xspace}
\newcommand{\fixedbetafermilsvalues}{1.10^{+0.26}_{-0.31}}
\newcommand{\fixedbetafermils}{\beta_\mathrm{LSP}=\fixedbetafermilsvalues\xspace}
\newcommand{\fixedbetafermimfvalues}{1.20^{+0.21}_{-0.23}}
\newcommand{\fixedbetafermimf}{\beta_\mathrm{MFVF}=\fixedbetafermimfvalues\xspace}
\newcommand{\fixedbetafermimeanvalue}{1.20^{+0.21}_{-0.23}}
\newcommand{\fixedbetafermimean}{\beta_\textrm{HE}=\fixedbetafermimeanvalue\xspace}

\newcommand{\betahesssfvalues}{1.00^{+0.24}_{-0.07}}
\newcommand{\betahesssf}{\beta_\mathrm{SF}=\betahesssfvalues\xspace}
\newcommand{\betahesslsvalues}{1.10^{+0.28}_{-0.09}}
\newcommand{\betahessls}{\beta_\mathrm{LSP}=\betahesslsvalues\xspace}
\newcommand{\betahessmfvalues}{1.10^{+0.23}_{-0.06}}
\newcommand{\betahessmf}{\beta_\mathrm{MFVF}=\betahessmfvalues\xspace}
\newcommand{\fminhesssfvalues}{-3.80^{+1.61}_{-0.18}}
\newcommand{\fminhesssflimit}{<-2.19}
\newcommand{\fminhesssf}{\mathrm{log}_{10}\left(f_\mathrm{min,SF}/\mathrm{d}^{-1}\right)=\fminhesssfvalues\fminhesssflimit\xspace}
\newcommand{\fminhesslsvalues}{-3.80^{+1.12}_{-0.31}}
\newcommand{\fminhesslslimit}{<-2.68}
\newcommand{\fminhessls}{\mathrm{log}_{10}\left(f_\mathrm{min,LSP}/\mathrm{d}^{-1}\right)=\fminhesslsvalues\fminhesslslimit\xspace}
\newcommand{\fminhessmfvalues}{-3.60^{+1.41}_{-0.34}}
\newcommand{\fminhessmf}{\mathrm{log}_{10}\left(f_\mathrm{min,MFVF}/\mathrm{d}^{-1}\right)=\fminhessmfvalues\xspace}

\newcommand{\betafermisfvalues}{1.20^{+0.44}_{-0.11}}
\newcommand{\betafermisf}{\beta_\mathrm{SF}=\betafermisfvalues\xspace}
\newcommand{\betafermilsvalues}{1.10^{+0.46}_{-0.11}}
\newcommand{\betafermils}{\beta_\mathrm{LSP}=\betafermilsvalues\xspace}
\newcommand{\betafermimfvalues}{1.30^{+0.54}_{-0.08}}
\newcommand{\betafermimf}{\beta_\mathrm{MFVF}=\betafermimfvalues\xspace}
\newcommand{\fminfermisfvalues}{-4.20^{+1.40}_{-0.12}}
\newcommand{\fminfermisflimit}{<-2.80}
\newcommand{\fminfermisf}{\mathrm{log}_{10}\left(f_\mathrm{min,SF}/\mathrm{d}^{-1}\right)=\fminfermisfvalues\fminfermisflimit\xspace}
\newcommand{\fminfermilsvalues}{-4.20^{+1.42}_{-0.12}}
\newcommand{\fminfermilslimit}{<-2.78}
\newcommand{\fminfermils}{\mathrm{log}_{10}\left(f_\mathrm{min,LSP}/\mathrm{d}^{-1}\right)=\fminfermilsvalues\xspace\fminfermilslimit}
\newcommand{\fminfermimfvalues}{-3.40^{+0.74}_{-0.12}}
\newcommand{\fminfermimf}{\mathrm{log}_{10}\left(f_\mathrm{min,MFVF}/\mathrm{d}^{-1}\right)=\fminfermimfvalues\xspace}

\begin{document}

\title{Characterizing the $\gamma$-ray long-term
variability of \Pk with \He and \FermiLAT} 

\authorrunning{The H.E.S.S. Collaboration}
\titlerunning{$\gamma$-ray long-term variability of \Pk}

\author{\small H.E.S.S. Collaboration
\and H.~Abdalla \inst{1}
\and A.~Abramowski \inst{2}
\and F.~Aharonian \inst{3,4,5}
\and F.~Ait Benkhali \inst{3}
\and A.G.~Akhperjanian\protect\footnotemark[2] \inst{6,5} 
\and T.~Andersson \inst{10}
\and E.O.~Ang\"uner \inst{7}
\and M.~Arrieta \inst{15}
\and P.~Aubert \inst{24}
% ...
}

\institute{
Centre for Space Research, North-West University, Potchefstroom 2520, South Africa \and 
Universit\"at Hamburg, Institut f\"ur Experimentalphysik, Luruper Chaussee 149, D 22761 Hamburg, Germany \and 
Max-Planck-Institut f\"ur Kernphysik, P.O. Box 103980, D 69029 Heidelberg, Germany \and 
Dublin Institute for Advanced Studies, 31 Fitzwilliam Place, Dublin 2, Ireland \and 
National Academy of Sciences of the Republic of Armenia,  Marshall Baghramian Avenue, 24, 0019 Yerevan, Republic of Armenia  \and
Yerevan Physics Institute, 2 Alikhanian Brothers St., 375036 Yerevan, Armenia \and
Institut f\"ur Physik, Humboldt-Universit\"at zu Berlin, Newtonstr. 15, D 12489 Berlin, Germany \and
University of Namibia, Department of Physics, Private Bag 13301, Windhoek, Namibia \and
GRAPPA, Anton Pannekoek Institute for Astronomy, University of Amsterdam,  Science Park 904, 1098 XH Amsterdam, The Netherlands \and
Department of Physics and Electrical Engineering, Linnaeus University,  351 95 V\"axj\"o, Sweden \and
Institut f\"ur Theoretische Physik, Lehrstuhl IV: Weltraum und Astrophysik, Ruhr-Universit\"at Bochum, D 44780 Bochum, Germany \and
GRAPPA, Anton Pannekoek Institute for Astronomy and Institute of High-Energy Physics, University of Amsterdam,  Science Park 904, 1098 XH Amsterdam, The Netherlands \and
Institut f\"ur Astro- und Teilchenphysik, Leopold-Franzens-Universit\"at Innsbruck, A-6020 Innsbruck, Austria \and
School of Physical Sciences, University of Adelaide, Adelaide 5005, Australia \and
LUTH, Observatoire de Paris, PSL Research University, CNRS, Universit\'e Paris Diderot, 5 Place Jules Janssen, 92190 Meudon, France \and
Sorbonne Universit\'es, UPMC Universit\'e Paris 06, Universit\'e Paris Diderot, Sorbonne Paris Cit\'e, CNRS, Laboratoire de Physique Nucl\'eaire et de Hautes Energies (LPNHE), 4 place Jussieu, F-75252, Paris Cedex 5, France \and
Laboratoire Univers et Particules de Montpellier, Universit\'e Montpellier, CNRS/IN2P3,  CC 72, Place Eug\`ene Bataillon, F-34095 Montpellier Cedex 5, France \and
DSM/Irfu, CEA Saclay, F-91191 Gif-Sur-Yvette Cedex, France \and
Astronomical Observatory, The University of Warsaw, Al. Ujazdowskie 4, 00-478 Warsaw, Poland \and
Aix Marseille Universit\'e, CNRS/IN2P3, CPPM UMR 7346,  13288 Marseille, France \and
Instytut Fizyki J\c{a}drowej PAN, ul. Radzikowskiego 152, 31-342 Krak{\'o}w, Poland \and
Funded by EU FP7 Marie Curie, grant agreement No. PIEF-GA-2012-332350,  \and
School of Physics, University of the Witwatersrand, 1 Jan Smuts Avenue, Braamfontein, Johannesburg, 2050 South Africa \and
Laboratoire d'Annecy-le-Vieux de Physique des Particules, Universit\'{e} Savoie Mont-Blanc, CNRS/IN2P3, F-74941 Annecy-le-Vieux, France
% ...
}

\offprints{H.E.S.S.~collaboration,
\protect\\\email{\href{mailto:contact.hess@hess-experiment.eu}{contact.hess@hess-experiment.eu}};
\protect\\\protect\footnotemark[1] Corresponding authors
\protect\\\protect\footnotemark[2] Deceased
}


\date{Received 28-07-2016 /
Accepted 29-09-2016}


\abstract{Studying the temporal variability of BL Lac objects 
at the highest energies provides unique insights into the extreme physical processes 
occurring in relativistic jets and in the vicinity of super-massive black holes.  To this 
end, the long-term variability of the BL Lac object \Pk is analyzed in the 
high (HE, $100\,$MeV$<E<300\,$GeV) and very high energy (VHE, $E>200\,$GeV) 
$\gamma$-ray domain. Over the course of $\sim 9\,$yr of \He observations the VHE 
light curve in the quiescent state is consistent with a log-normal behavior. The VHE
variability in this state is well described by flicker noise (power-spectral-density index
$\fixedbetahessmean$) on time scales larger than one day.  An analysis of 
$\sim5.5$ yr of HE \FermiLAT data gives consistent results ($\fixedbetafermimean$, on time scales
larger than 10 days) compatible with the VHE findings. The HE and VHE power spectral 
densities show a scale invariance across the probed time ranges. A direct linear correlation 
between the VHE and HE fluxes could neither be excluded nor firmly established. 
These long-term-variability
properties are discussed and compared to the red noise behavior ($\beta\sim2$) seen 
on shorter time scales during VHE-flaring states. The difference in power spectral noise
behavior at VHE energies during quiescent and flaring states provides
evidence that these states are influenced by different physical
processes, while the compatibility of the HE and VHE long-term results
is suggestive of a common physical link as it might be introduced by
an underlying jet-disk connection.}


\maketitle


\section{Introduction}

\noindent One of the most striking properties of BL Lacertae
objects is their variability across the electromagnetic
spectrum from radio to $\gamma$ rays and accros the temporal spectrum
from minutes to years.  In the current standard paradigm of active galactic nuclei
\citep[e.g.,][]{1995PASP..107..803U} the observed nonthermal
emission is produced in two-sided collimated, relativistic plasma
outflows (jets) closely aligned with the line of sight, so that
the intrinsic emission appears enhanced because of
Doppler-boosting effects. The jets are powered by a central
engine consisting of a supermassive black hole surrounded by an
accretion disk. Characterizing the temporal variability provides
one of the key diagnostics for the physical conditions in these
systems, for example the jet-disk connection, location of the
emitting region, or dominant emission processes.

\noindent The BL Lac \Pk \citep[redshift
$z=0.116$;][]{1993ApJ...411L..63F} has been observed with the High 
Energy Stereoscopic System \citep[H.E.S.S.;][]{2004NewAR..48..331H} 
at very high-energy $\gamma$-rays (VHE; $E>200$\,GeV) since 2002 
\citep[e.g.,][and references therein]{2010A&A...520A..83H}.  The source 
underwent an extreme VHE flux outburst in July/August 2006 with peak 
fluxes exceeding the average flux level of the long-term emission by a 
factor of $\sim$\,$100$, during which it showed rapid variability on timescales 
as short as 3\,min \citep{2007ApJ...664L..71A}. The stochastic VHE variability 
during this flaring state has been characterized as power-law noise 
($\propto f^{-\beta}$, where $f$ is the frequency) with an index $\beta \sim2$.  
A detailed analysis of the VHE data from 2005-2007 revealed that the run-by-run
light curve of \Pk follows a skewed flux distribution, 
which is well represented by two superposed lognormal distributions 
\citep[][Fig.~3]{2010A&A...520A..83H}. 
Excluding the flare data, the flux distribution satisfies a simple lognormal 
distribution. This provides evidence that the source switches from a 
quiescent VHE state with minimal activity to a flaring state, and the flux
distribution in each state follows a lognormal distribution.
Lognormal behavior was first established for accreting Galactic sources
such as X-ray binaries by \cite{2001MNRAS.323L..26U}, linking such a behavior 
to the underlying accretion process.
In a lognormal process, the fluctuations of 
the flux are on average proportional, or at least correlated, to the flux itself, 
ruling out additive processes in favor of multiplicative processes. 
In the case of blazars, a lognormal behavior could thus mark the influence of 
the accretion disk on the jet \citep[e.g.,][]{2009A&A...503..797G,2010LNP...794..203M}.
Cascade-like events are an example of multiplicative lognormal processes. 
Density fluctuations in the accretion disk provide one possible realization. If 
damping is negligible, these fluctuations can propagate inward and couple 
together to produce a multiplicative behavior. If this is efficiently transmitted 
to the jet, the $\gamma$-ray emission could be modulated accordingly.

\noindent In Sect.~2 we present VHE data from $9$ years of \He observations of 
\Pk in the quiescent state, and partially contemporaneous HE data from $5.5$ years of 
observations from the Large Area Telescope (\FermiLAT), are presented in Sec.~2.  
In Sect.~3, a detailed time-series analysis is performed. First, the light curves 
are tested for a lognormal behavior. Then, their variability is
characterized as power-law noise with a forward folding method with simulated
light curves, taking the sampling of the data,
described in \citet{2011A&A...531A.123K} into account. And, finally, the VHE and HE
emissions are also analyzed for a possible direct 
linear correlation. This is the first time that such an extended
analysis is performed on the VHE $\gamma$-ray emission of a BL
Lac object on timescales as long as nine years. In Sect.~4 the results 
of the two energy ranges are compared and their implications on the
physical properties of \Pk are discussed.


\section{Observations and analysis}\label{Sec:Data}

\begin{figure*}
\centering
\includegraphics[width=1.0\textwidth]{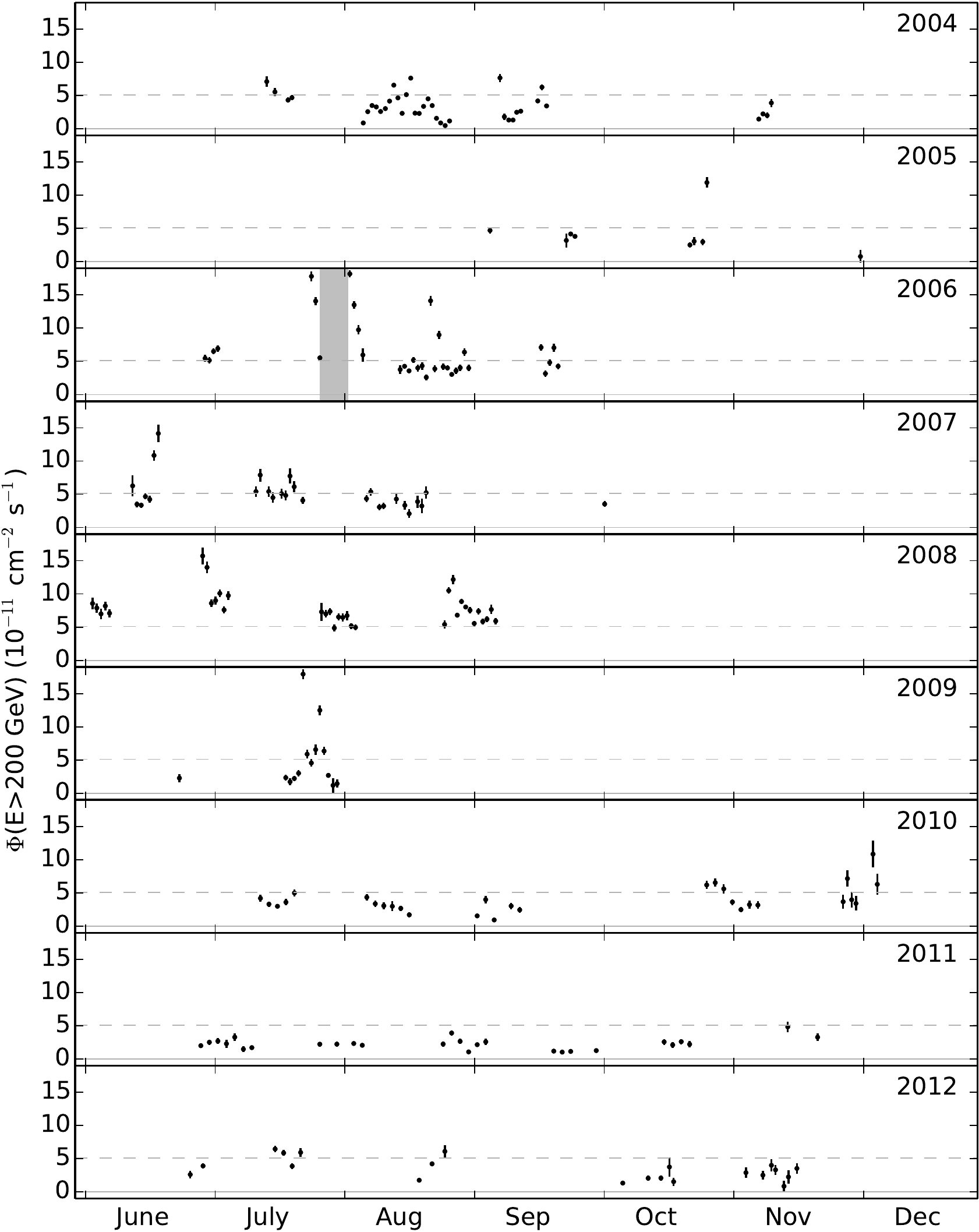}
\caption{\He light curve of nightly fluxes above 200\,GeV excluding the high state in July/August 2006 (gray shaded area). The gray dashed horizontal line indicates the average flux of the quiescent state.  } 
\label{Fig:HESSLightcurve} 
\end{figure*}
 
\paragraph{\He (VHE):}
H.E.S.S. is an array of five
Imaging Atmospheric Cherenkov Telescopes (IACTs). The first phase of 
\He began in 2003 with four 12-meter telescopes giving an energy threshold
$\sim$\,$100$ GeV. In 2012, a fifth 28-meter telescope was added
to the array, reducing the energy threshold to $\sim$\,$50$ GeV.

\noindent The present analysis is based on VHE data taken with the completed \He 
Phase\hbox{-}I between MJD $53200$ (14 July 2004) and $56246$ 
(15 November 2012), using three or four telescopes. The high flux state from 27 July to 
8 August 2006, with an average flux of 
$(75.2 \pm 0.8) \times 10^{-11}$ cm$^{-2}$ s$^{-1}$ 
above $200$ GeV is excluded from the data set. 
The remaining data constitutes the basis for the time-series 
analysis, and is referred to as the quiescent data set in the following.

\noindent In total, about $328\,$h of data passed standard quality cuts 
as defined in \citet{2006A&A...457..899A}, with a mean zenith angle of 
$21^\circ$ resulting in an average energy threshold of $178\,$GeV.  The data set 
has been analyzed with the Model analysis chain using standard 
cuts \citep{2009APh....32..231D} above $200$ GeV.

\noindent  The total detection significance in the quiescent data set 
corresponds to $341\,\sigma$. The light curve of nightly fluxes is calculated 
assuming a log-parabolic energy spectrum, 
\begin{equation}
\centering
 {\rm d}N/{\rm d}E \propto E^{-a - b \log E},
 \label{eq:logparabola}
 \end{equation}
with $a$ and $b$ corresponding to the best-fit power-law index and
curvature index, respectively. In order to take indications of a
spectral variability in the VHE domain during the quiescent state
\citep{2010A&A...520A..83H} into account, the nightly fluxes are
derived with a separate log-parabola fit of the spectrum for each
year.

\noindent The values for $a$ and $b$ are summarized in
Table~\ref{table:spectrumvalue}; the average values are $\overline{a}
= 3.209$ and $\overline{b} = 0.164$.  The unbiased sample variance for
the first parameter
\begin{align}s_a^2=\frac{1}{n-1}\sum_{i=1}^n\left(a_i-\overline{a}\right)^2=0.017\end{align}
is larger than the expected variance $\overline{\sigma_{a}^2}=0.005$ because
of the uncertainties $\sigma_{a}$ on the individual best-fit values.
The same is true for the second parameter with $s_b^2=0.114$ and
$\overline{\sigma_{b}^2}=0.005$. This could be indicative of a
variable spectrum.

\begin{table}
\caption{Values of the log-parabola parameters of the spectrum (Eq.~\ref{eq:logparabola}) of \Pk used to derive the light curve.}            
\label{table:spectrumvalue} 
\centering            
\begin{tabular}{l c c}       
\hline \hline                  
Year & $a\pm\sigma_{a}$ & $b\pm\sigma_{b}$ \\ 
\hline 
2004 & $2.95 \pm 0.03$ & $0.37 \pm  0.03$ \\ 
2005 & $3.27 \pm 0.12$ & $0.25 \pm  0.12$ \\ 
2006 & $3.27 \pm 0.04$ & $0.24 \pm  0.04$ \\ 
2007 & $3.38 \pm 0.08$ & $-0.03 \pm 0.07$ \\
2008 & $3.28 \pm 0.04$ & $0.12 \pm  0.03$ \\ 
2009 & $3.14 \pm 0.08$ & $0.18 \pm  0.07$ \\ 
2010 & $3.24 \pm 0.08$ & $0.10 \pm  0.07 $ \\ 
2011 & $3.08 \pm 0.10$ & $0.13 \pm 0.08$ \\ 
2012 & $3.27 \pm 0.01$ & $0.12 \pm  0.09$ \\
\hline                
\end{tabular}
\end{table}

\noindent Variations within a season, however, are unlikely to affect the analysis
presented here as the inferred small changes in the spectral
parameters only result in small changes in the integral flux.

\noindent The resulting quiescent light curve has an average flux of
$(5.10 \pm 0.41) \times10^{-11}\,\mathrm{cm}^{-2}\,\mathrm{s}^{-1}$
above 200 GeV and is shown in Fig.~\ref{Fig:HESSLightcurve}.
It is characterized by a fractional root mean square variability 
\citep{2003MNRAS.345.1271V}
\begin{equation}
F_{\rm var} = {\sqrt{S_\Phi^2- \overline{\sigma_{\rm err}^2}} \over
\overline{\Phi}} = 0.66 \pm 0.01,
\label{eq:fvar}
\end{equation}
where $\overline{\Phi}$ is the mean flux, $S_\Phi^2$ is the variance of the
fluxes and
$\overline{\sigma_{\rm err}^2}$ is the contribution to the variance caused by
the measurement errors. 
The analysis results were
cross-checked using different analysis methods and calibration
chains \citep[e.g.,][]{2006A&A...457..899A}, yielding compatible results.

\paragraph{\FermiLAT (HE):} 
\begin{figure}
\centering
\includegraphics[width=0.45\textwidth]{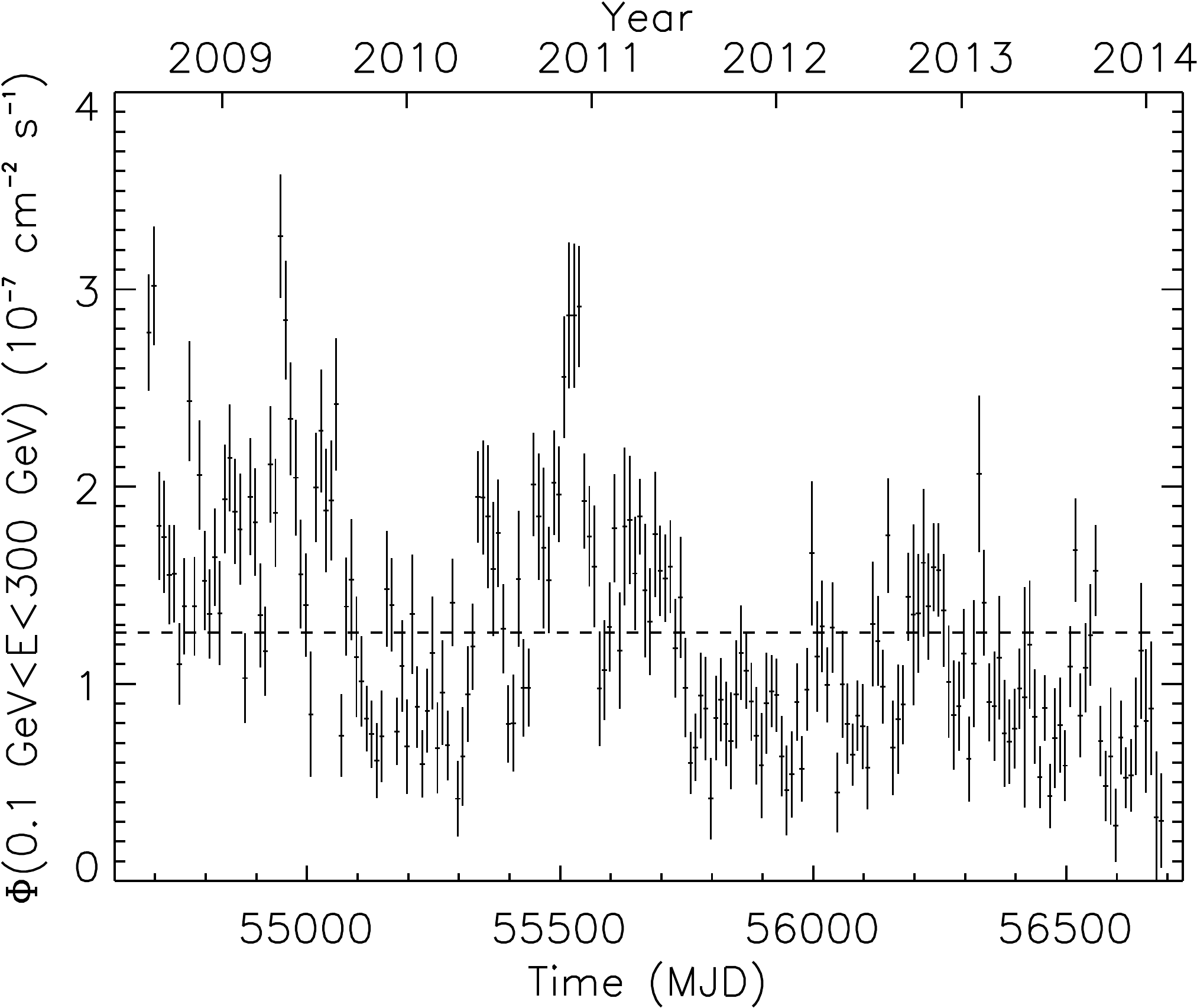}
\caption{Light curve of the integral fluxes between $0.1$ and $300\,$GeV in bins of ten days measured with \FermiLAT.  The dashed line indicates the average flux.} 
\label{Fig:FermiLightcurve}
\end{figure}

The \FermiLAT \citep{2009ApJ...697.1071A} on board the \Fermi
satellite is a pair-conversion telescope designed to detect
HE $\gamma$-rays in the energy range from below
$20\,$MeV to more than $300\,$GeV.

\noindent The set of observational data for \Pk used here covers
about $5.5$ years, from MJD $54688$ (10 August 2008) to $56688$ 
(31 January 2014). Events are selected between 100 MeV and 300 GeV in 
a region of interest (ROI) of $15\degree$ around \Pk. The detector is 
described by the P7REP\_SOURCE\_V15 instrument response function\footnote{\url{http://fermi.gsfc.nasa.gov/ssc/data/analysis/documentation/Pass7REP\_usage.html}}.  
The HE light curve is obtained with the tool Enrico \citep{2013arXiv1307.4534S} 
using the \Fermi Science Tools v9r32p5\footnote{cf. the \Fermi Science Support
Center website \url{http://fermi.gsfc.nasa.gov/ssc/}}, with a 10-day binning 
to ensure enough statistics in each bin. 
The prefactor of the diffuse galactic
background (gll\_iem\_v05) and the normalization of the isotropic
diffuse emission (iso\_source\_v05) are left free to vary in the
likelihood fit.

\noindent All sources from the third LAT source catalog
\citep[3FGL;][]{2015ApJS..218...23A} within $15\degree$ of \Pk
are included in the model to ensure a good background modeling. 
The HE spectra of \Pk and all sources
within $3\degree$ are fitted following the spectral shape of the 3FGL catalog. The spectra of \Pk, 
3FGL~J2151.6-2744 and 3FGL~J2159.2-2841 are modeled with a simple power law, 
while for 3FGL J2151.8-3025 a log-parabolic shape is assumed.
The indices and pre-factors are left free. All other components are fixed 
to the values in the 3FGL catalog. \Pk is detected with a significance of 
$156\,\sigma$ with a spectral index of $1.83 \pm 0.01$. The photon 
counts of \Pk are integrated into $201$ bins of ten days.
The resulting light curve has an average flux of
$1.20 \pm 0.03 \times10^{-7}\,\mathrm{cm}^{-2}\,\mathrm{s}^{-1}$ 
between 100 MeV and 300 GeV with a 
$F_{\rm var} = 0.41 \pm 0.02$ (Eq.~\ref{eq:fvar}) and is shown in 
Fig.~\ref{Fig:FermiLightcurve}. 


\section{Time-series analysis}\label{Sec:Results}

\noindent To characterize the long-term variability of \Pk, the HE and VHE data sets are 
analyzed with respect to lognormality and power-law noise. 


\noindent {\it Lognormality:} A possible lognormal behavior of \Pk is 
investigated by examining the distribution of the fluxes and studying the correlation 
between the flux levels and the intrinsic variability.
\noindent The flux and log-flux distributions of each light curve are fitted by a Gaussian 
using a $\chi^2$ fit. The goodness of the fits are summarized in Table \ref{table:histo}. 
In both cases, a Gaussian fits the log-flux distribution better than the flux 
distribution, with a significance level of $\sigma > 5$ for the VHE and $\sigma > 2$ 
for the HE data, respectively. For the VHE data the probability representing the 
goodness of fit is about $10^4$ times higher for the log-flux distribution when 
compared to the flux distribution. For the HE data this ratio is on the order of 
$10$. Thus, while the HE $\gamma$-ray flux of \Pk in this approach shows only 
an indication, the VHE $\gamma$-ray flux data provide evidence for a lognormal 
behavior.

\begin{table}
\caption{Values of the reduced $\chi^2$ and associated probability for the Gaussian fits of the flux and log flux distributions for each light curve. The parameter $\sigma$ is the significance level on which a lognormal distribution is preferred to a normal distribution.}            
\label{table:histo} 
\centering            
\begin{tabular}{l c c c c c}       
\hline \hline 
& \multicolumn{2}{c}{$\Phi$} & \multicolumn{2}{c}{log $\Phi$} & \\ 
& $\chi^2/$d.o.f. & Prob & $\chi^2/$d.o.f. & Prob & $\sigma$ \\ 
\hline \He & 50.8/17 & $10^{-5}$ & 11.9/13 & 0.54 & 5.39 \\ 
\Fermi & 21.6/12 & 0.04 & 15.0/11 & 0.18 & 2.57 \\ 
\hline            
\end{tabular}
\end{table}

\begin{figure*}
\centering
\includegraphics[width=0.49\textwidth]{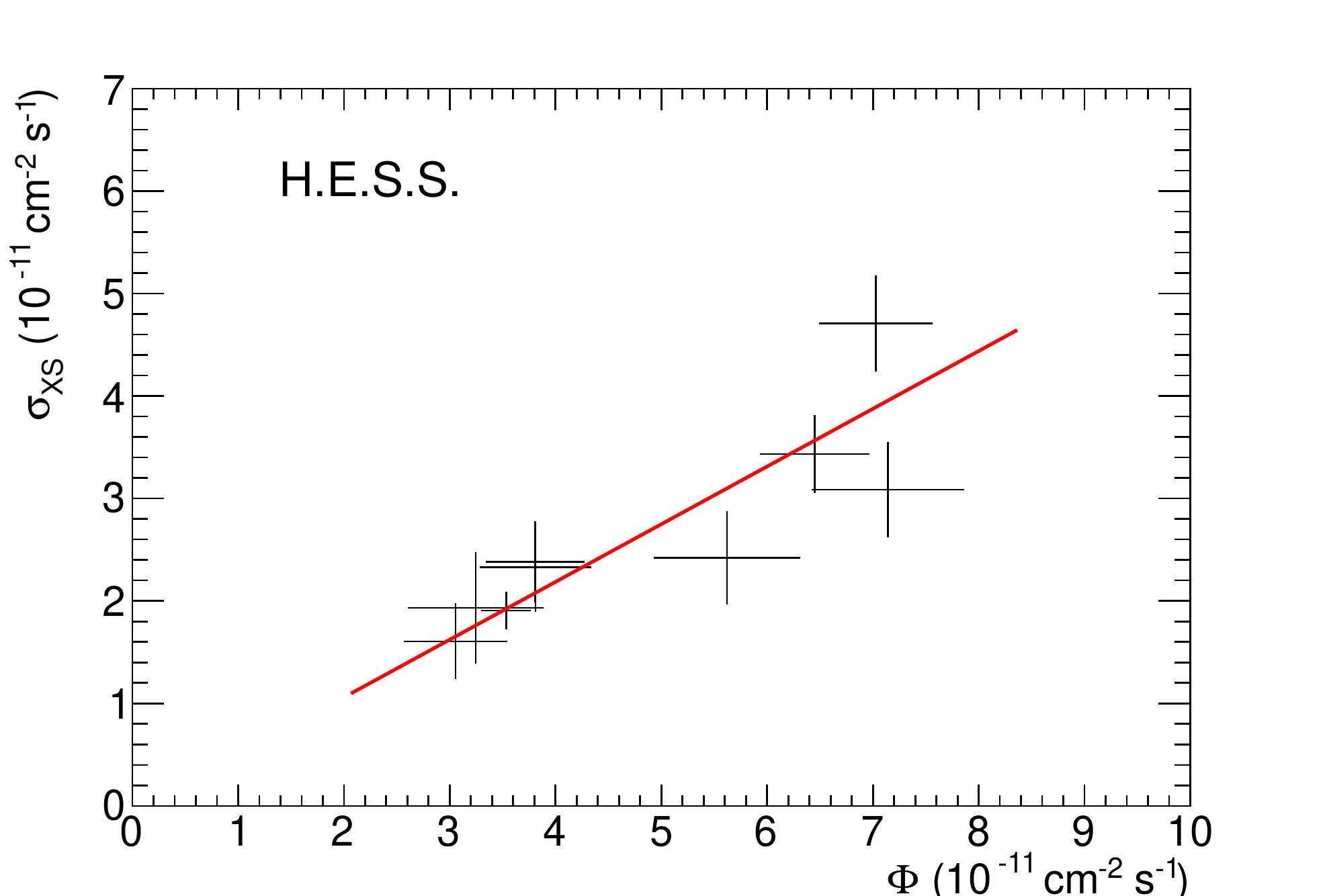}
\includegraphics[width=0.49\textwidth]{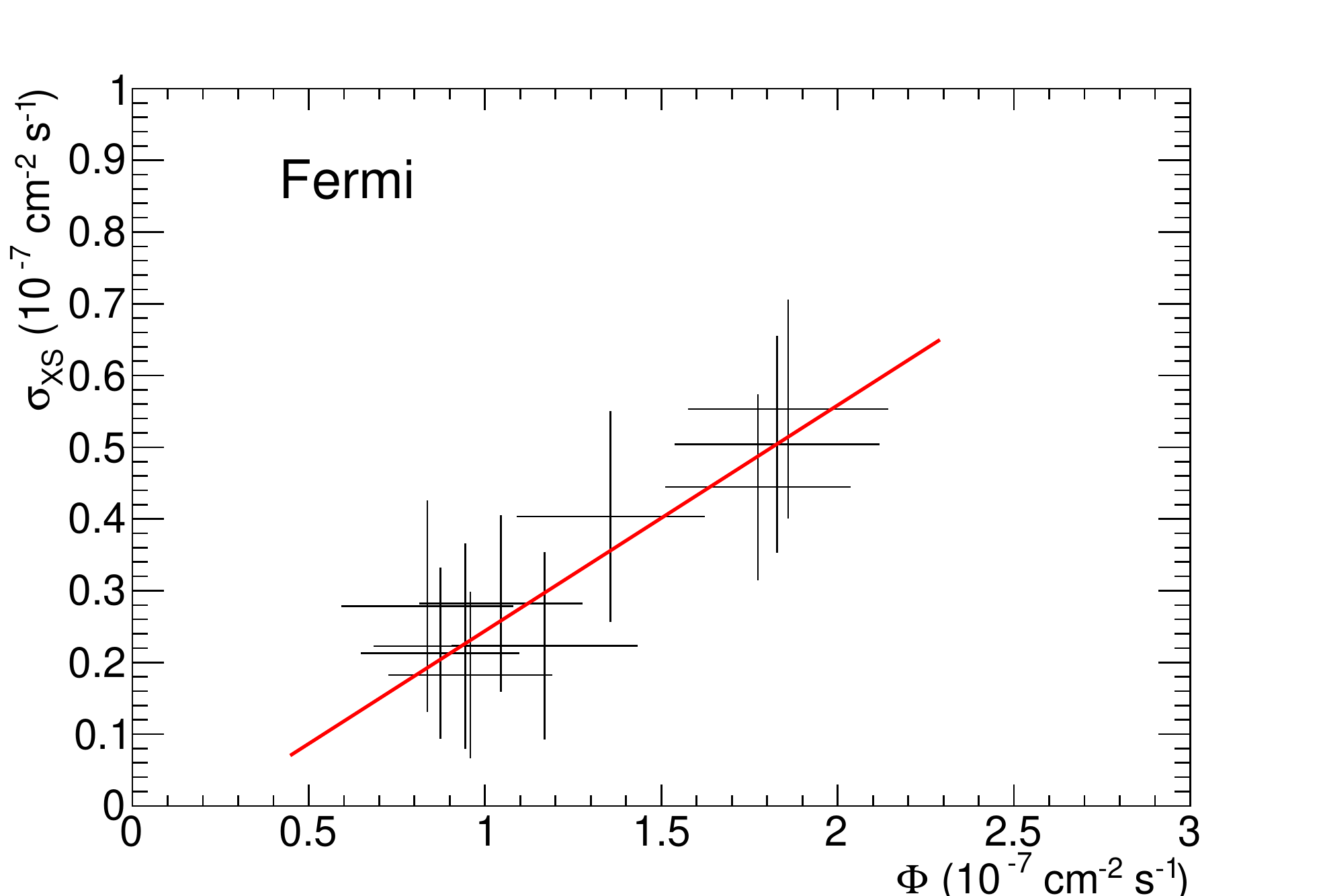}
\caption{Scatter plot of the excess RMS and the average flux for the H.E.S.S. (left) and \FermiLAT (right) data. Each flux and excess rms values are computed using at least 20 adjacent light curve points. A linear fit is shown in red.} 
\label{plot:excess} 
\end{figure*}

\noindent In addition, the variability-flux relation, estimated by the excess RMS (root mean square;
Eq.~\ref{eq:sigmaexcess}), is investigated for a possible correlation \citep{2005MNRAS.359..345U}. 
The excess RMS estimates the intrinsic variability 
of a time series by subtracting the contribution of the measurement errors. It is 
defined as 
\begin{equation}
\sigma_{\rm XS} = \sqrt{S^2- \overline{\sigma_{\rm err}^2}}
\label{eq:sigmaexcess}
\end{equation}
where $S^2$ is the variance and $ \overline{\sigma_{\rm err}^2}$ the mean square 
of the statistical error of the data \citep{2003MNRAS.345.1271V}.  Here $\sigma_{\rm XS}$ is 
calculated for bins of the light curves each containing at least 20 light curve points. 
They are plotted versus the average fluxes $\overline{\Phi}$ of the corresponding 
bins for both light curves in Fig.~\ref{plot:excess}.

\noindent To test the possible correlation between the flux and its variability, the scatter plots 
are fitted by a constant as well as a linear ascending slope. The fit results are 
summarized in Table~\ref{table:excess}. The results reveal a preference greater
than $6 \sigma$ for the linear fit to the VHE data while HE data only show an
indication of linearity. To characterize this behavior beyond the fit of a linear correlation, 
the nonparametric correlation factor Spearman $\rho$ and the Kendall rank $\tau$,
which measure the ordering of the points, \citep{2004A&A...414.1091G}
are calculated. In all cases $\rho > 0.85$ and $\tau > 0.65$, meaning that 
$\sigma_{\rm XS}$ and $\overline{\Phi}$ show a strong correlation. This 
implies that the fluctuations of the flux are in fact correlated with the flux.

\begin{table}
\caption{Values of the reduced $\chi^2$ of the constant and linear fits of the scatter plots shown in Fig.~\ref{plot:excess} for each light curve with values for the significance $\sigma$, correlation factor $\rho$, and Kendall rank $\tau$.}            
\label{table:excess} 
\centering            
\begin{tabular}{l c c c c c}       
\hline \hline 
& constant & linear increase &  &  \\ 
& $\chi^2/$d.o.f. & $\chi^2/$d.o.f. & $\sigma$ \\ 
\hline H.E.S.S. & 5.78 & 0.89 & 6.33  \\ 
\Fermi & 0.936 & 0.107 & 2.75 \\ 
\hline 
\\
\hline \hline 
& $\rho$ & $\tau$ \\ 
\hline 
H.E.S.S. &$0.86 \pm 0.11$ & $0.78 \pm 0.26$  \\ 
\Fermi & $0.93 \pm 0.19$ & $0.69 \pm 0.25$\\ 
\hline            
\end{tabular}
\end{table}

\noindent The preceding analysis shows that the HE and VHE flux distributions 
of \Pk during the quiescent state are compatible with lognormal distributions.
When the two energy bands are compared, the lognormal behavior is much more 
evident in the VHE data set. In addition, for both energy bands the variability 
amplitude of the flux is correlated with the flux level, supporting the conclusion that 
lognormality is an intrinsic characteristic of the long-term $\gamma$-ray emission 
in \Pk. A similar result has also been reported for the VHE flaring state of \Pk in 2006 
\citep{2010A&A...520A..83H}.

\noindent {\it Power-law noise:} The flux variability of active galaxies has frequently been 
characterized as power-law noise \citep[e.g.,][]{1993ApJ...414L..85L}. The power 
spectral density (PSD), i.e., the square modulus of the discrete Fourier transform
\citep[][]{1967JSV.....6...86P}, of such a light curve follows a power law $\propto 
f^{-\beta}$, where $f$ is the temporal frequency and usually $1 \simlt \beta \simlt 2$. 
The PSD reveals how the variability amplitude is distributed among the timescales. 
In a power-law noise light curve ($\beta > 0$) the flux variations on longer timescales 
dominate the variations on shorter timescales. The total variance of such a 
light curve tends to grow with its length. For $\beta=0$ the variability power is equal 
on each time cale, resulting in white noise, where the fluxes at all times are 
uncorrelated.

\noindent To study the variability characteristics, we first hypothesize that the light 
curves can be described by power-law noise with a lognormal behavior with
$\beta \ge 0$ as the only free parameter. Three different methods are then used 
to characterize the variability of the light curves: The Lomb-Scargle Periodogram 
(LSP), which is a method to approximate the PSD
\citep{1976Ap&SS..39..447L,1982ApJ...263..835S}; the first-order
Structure Function \citep[SF;][]{1985ApJ...296...46S} and the
Multiple Fragments Variance Function \citep[MFVF;][]{2011A&A...531A.123K},
which are both representations of the PSD in the time domain.

\noindent A forward-folding method with sets of $10^4$ simulated 
light curves for each value of $\beta$ and a maximum likelihood 
estimator are applied to estimate the best-fit parameter $\beta$, as 
described in \citet{2011A&A...531A.123K}. The light curves are simulated with
a lognormal behavior, so that the logarithm of the intrinsic light
curve $\Phi_{\rm intr}$ is power-law noise with a Gaussian behavior of mean $\mu$
and standard deviation $\sigma$. 
\begin{equation}
 \log (\Phi_{\rm intr}) \mapsto \mathcal{N}(\mu,\sigma) 
\end{equation}
The PSD of the simulated power-law noise is generated on a frequency range
$\left[(10T)^{-1},\,5\mathrm{d}^{-1}\right]$ where $T_\textrm{H.E.S.S.}=3047\,$d
and $T_{Fermi}=2000\,$d are the lengths of the measured light
curves, respectively.  The power-law noise is rebinned to ten days 
for the \FermiLAT analysis corresponding to the binning of the 
\Pk light-curve. The light curves are then downsampled to the real
observation times, the fluxes are rescaled to the average and
variance of the measured fluxes, and measurement errors are
simulated.
The parameter space is sampled with $\beta=0.0,0.1,\dots,3.0$.
The SFs, LSPs and MFVFs are equally binned in $\mathrm{log}_{10}$
scale with $50$, $20$, and $5$ bins per decade, respectively.

\noindent Owing to the binning, the smallest resolvable timescale
in the \FermiLAT light curve is $\tau_0=10\,$d and the
corresponding frequency is $f_\textrm{high}=(2\tau_0)^{-1}=(20\,\mathrm{d})^{-1}$.
For the \He light curve the smallest resolvable timescale is
$\tau_0=1\,$d corresponding to $f_\textrm{high}=(2\,\mathrm{d})^{-1}$. The LSPs
are characterized on a frequency range
$\left[10^{-4}\,\mathrm{d}^{-1},f_\textrm{high}\right]$, while the lowest timescale for the SFs and MFVFs
is $\tau_0$. The methods are also sensitive to variations that occur
on timescales larger than the lengths of the light curves and that
appear as long-term trends. It is therefore also useful to calculate 
the LSP on frequencies $<T^{-1}$ to reveal such information. 

\noindent The application of these methods to the \He light curve
in the quiescent state gives best-fit parameters 
$$\fixedbetahessls,\quad\fixedbetahesssf,\quad\fixedbetahessmf.$$
 
\noindent A goodness of fit test is applied where the maximum likelihood values 
of simulated LSPs, SFs and MFVFs, respectively, are used as test 
statistics. For this, new simulated sets of $10^4$ light curves of the best-fit parameters 
are analyzed with the likelihood estimator. The distribution of their 
maximum likelihood values is compared with the respective value of 
\Pk.  Assuming the model to be correct, the values should be compatible.  
The quantile that has a smaller likelihood than the \Pk data is used to estimate 
the p-value $p$. The smallest value of $p=11\%$ is found with 
the MFVF. The hypothesis that the data can be described by a single
power-law noise model is thus not rejected.

\noindent The MFVF gives the most precise value, which is taken 
here as the final value: 
$$\fixedbetahessmean.$$
The best-fit values of the other methods are compatible 
within the uncertainties.

\begin{figure*}[hb]
\centering
\begin{subfigure}[b]{0.49\textwidth} \caption{\He} \flushleft
\includegraphics[width=0.98\textwidth]{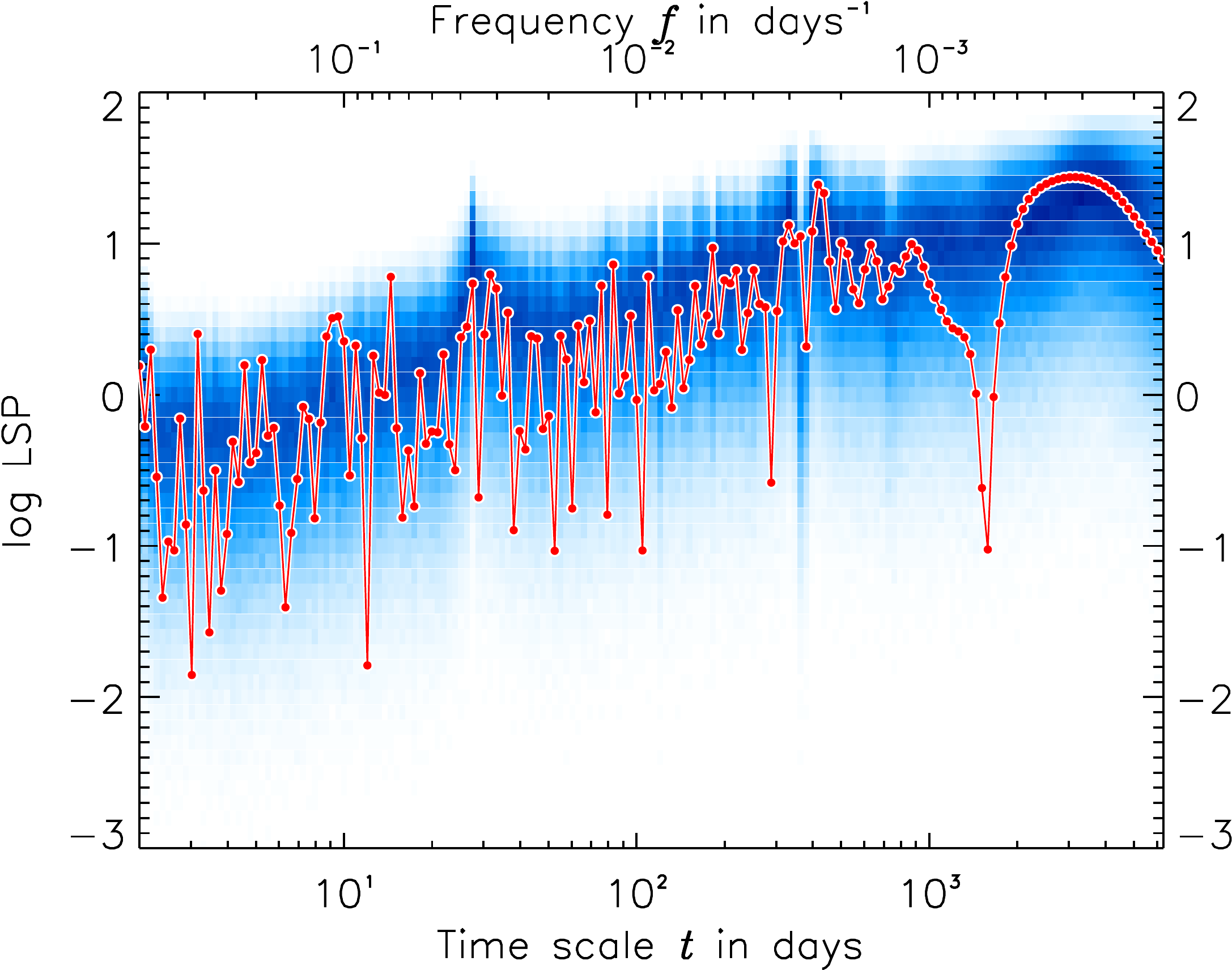}
\includegraphics[width=0.98\textwidth]{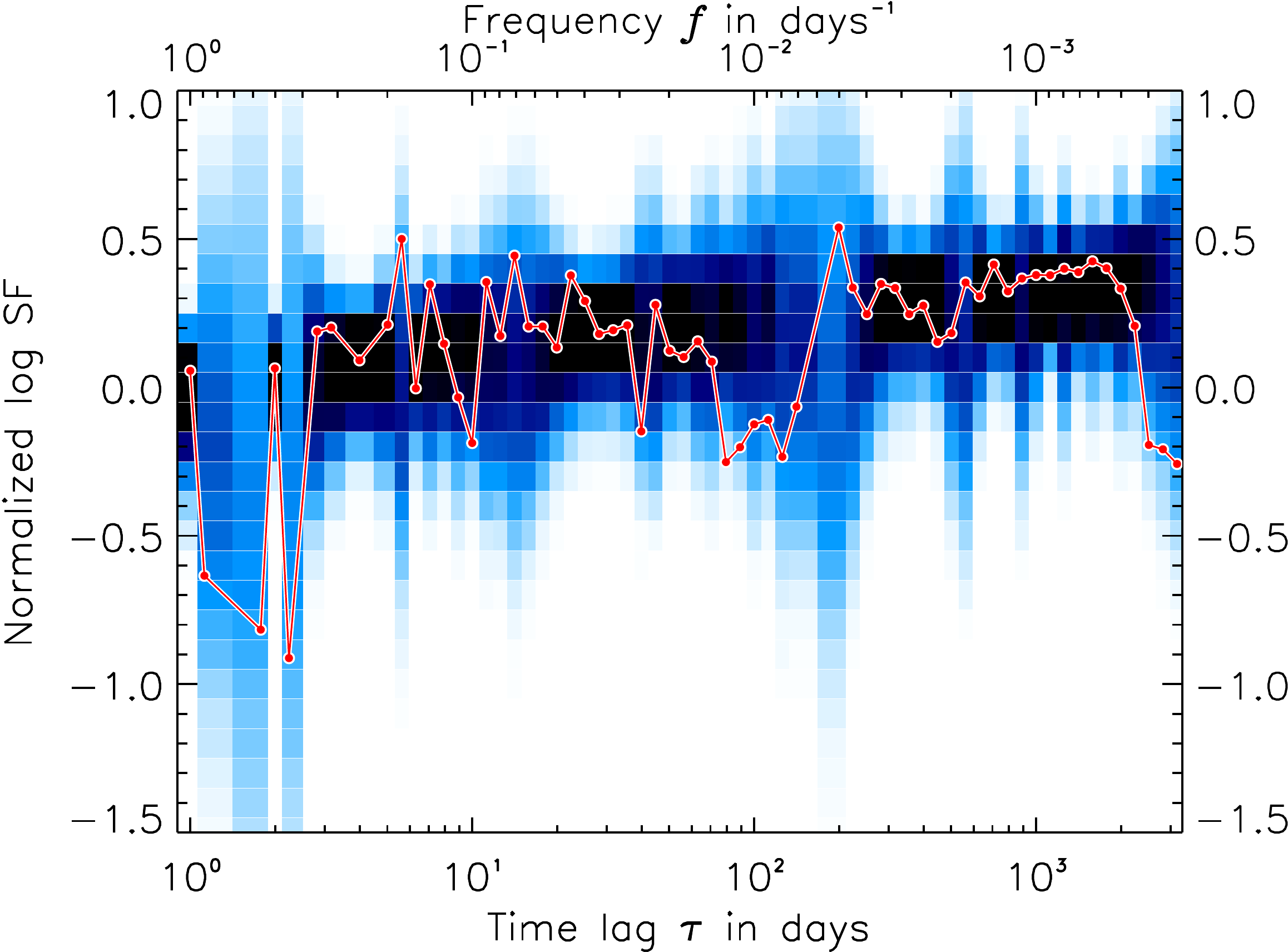}
\includegraphics[width=0.98\textwidth]{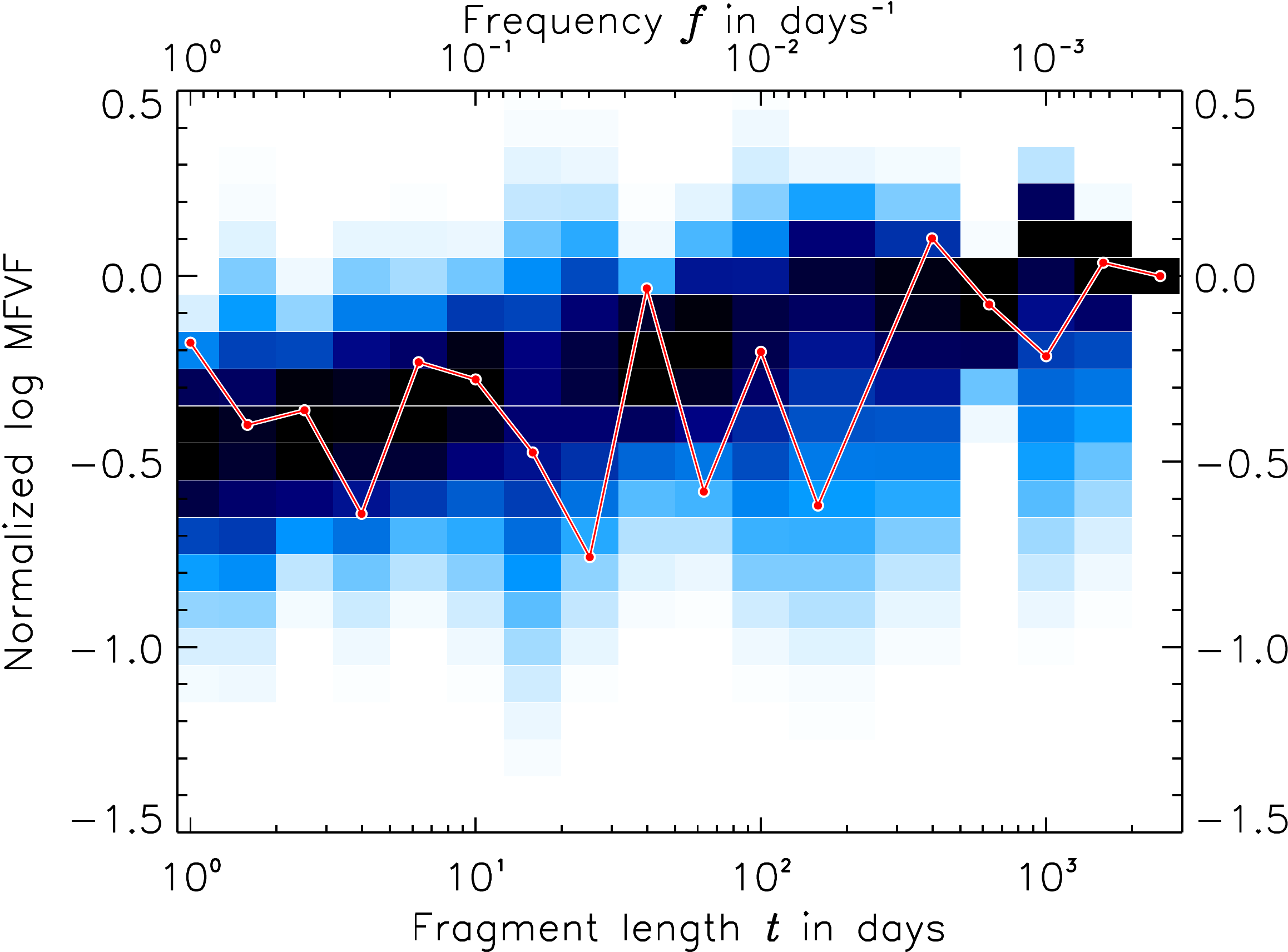}
\end{subfigure} \vline \begin{subfigure}[b]{0.49\textwidth}
\caption{\FermiLAT} \flushright
\includegraphics[width=0.98\textwidth]{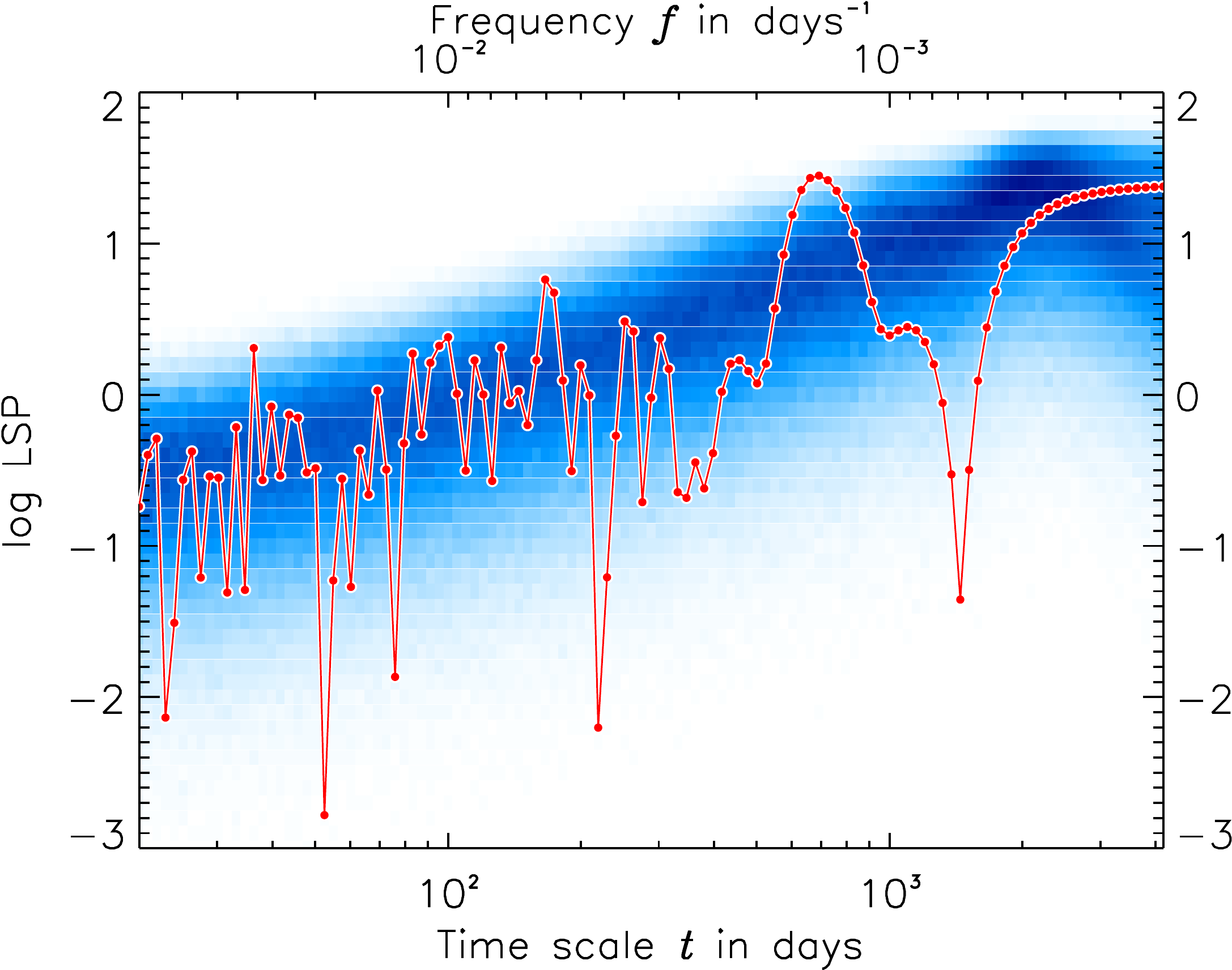}
\includegraphics[width=0.98\textwidth]{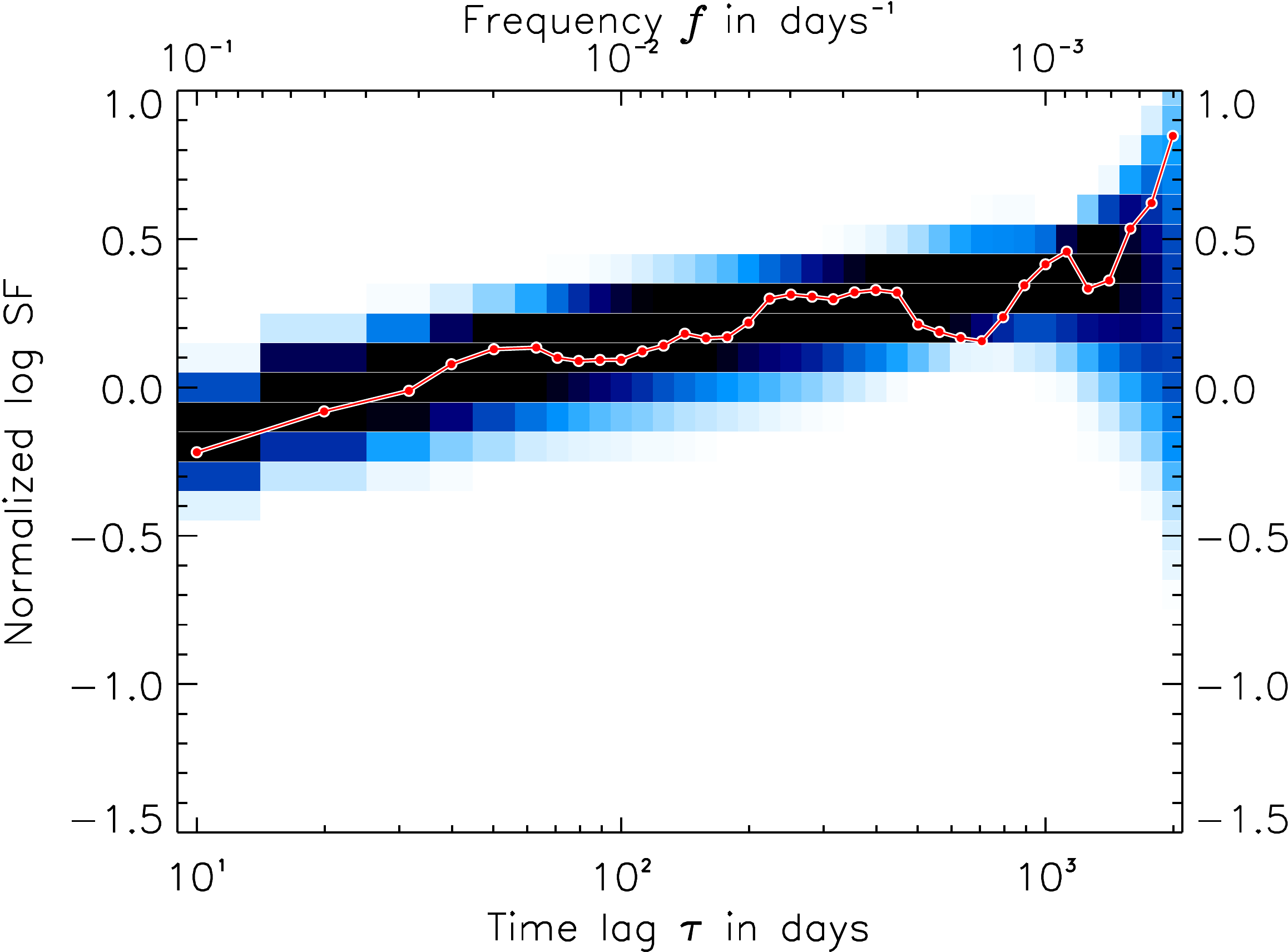}
\includegraphics[width=0.98\textwidth]{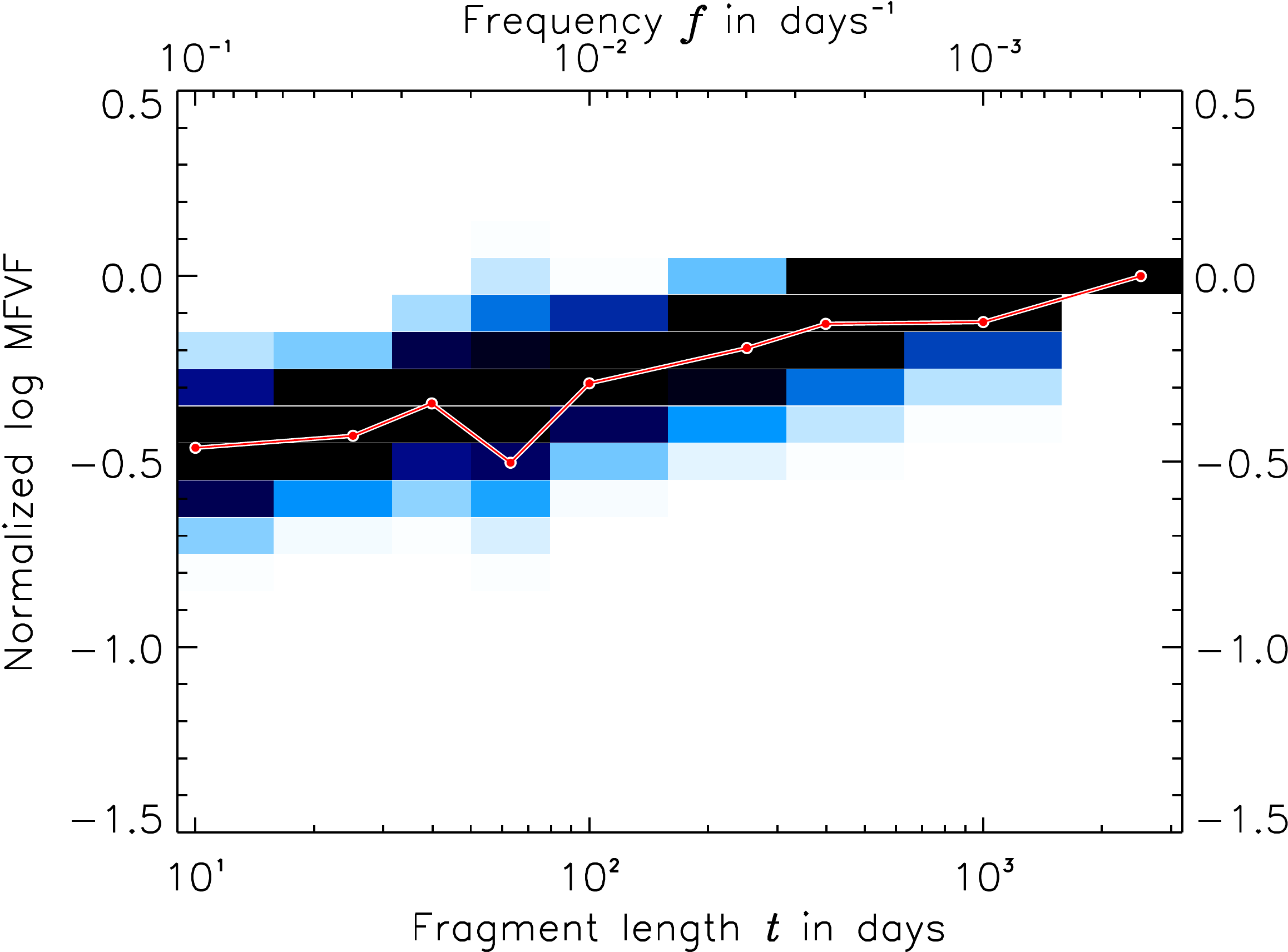}
\end{subfigure}
\includegraphics[width=0.98\textwidth]{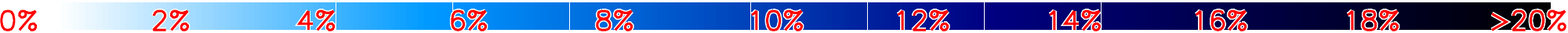}
\caption{LSPs, SFs, and MFVFs.  \emph{Left panel (a)}: The red solid lines are the LSP (top), SF (middle), and MFVF (bottom) for the \He quiescent light-curve. The best-fit PDFs of simulated LSP, SF, and MFVF values are represented by the blue histograms in color scale. \emph{Right panel} (b): Same plots for the \FermiLAT data.} 
\label{Fig:results} 
\end{figure*}

\begin{figure*}[hb]
\centering
\begin{subfigure}[b]{0.49\textwidth} \caption{\He} \flushleft
\includegraphics[width=0.98\textwidth]{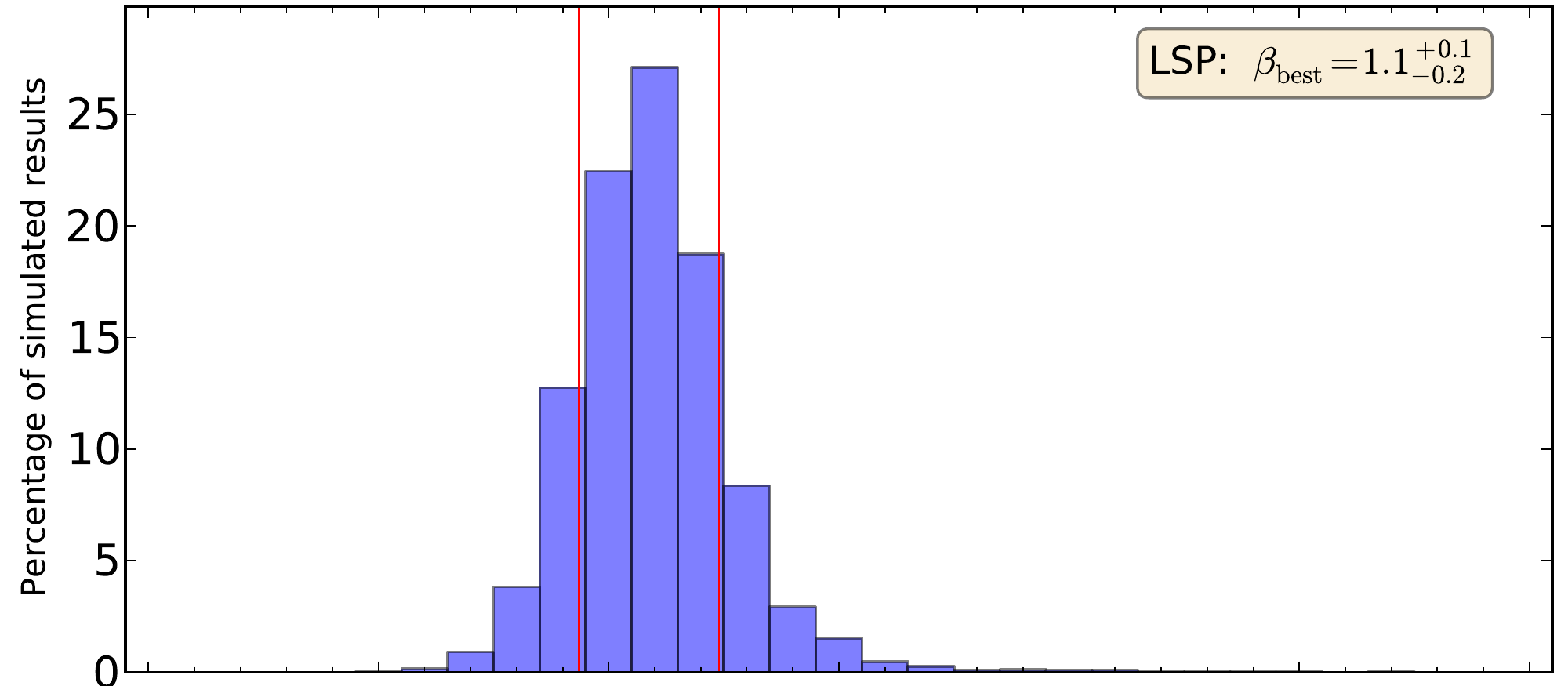}
\includegraphics[width=0.98\textwidth]{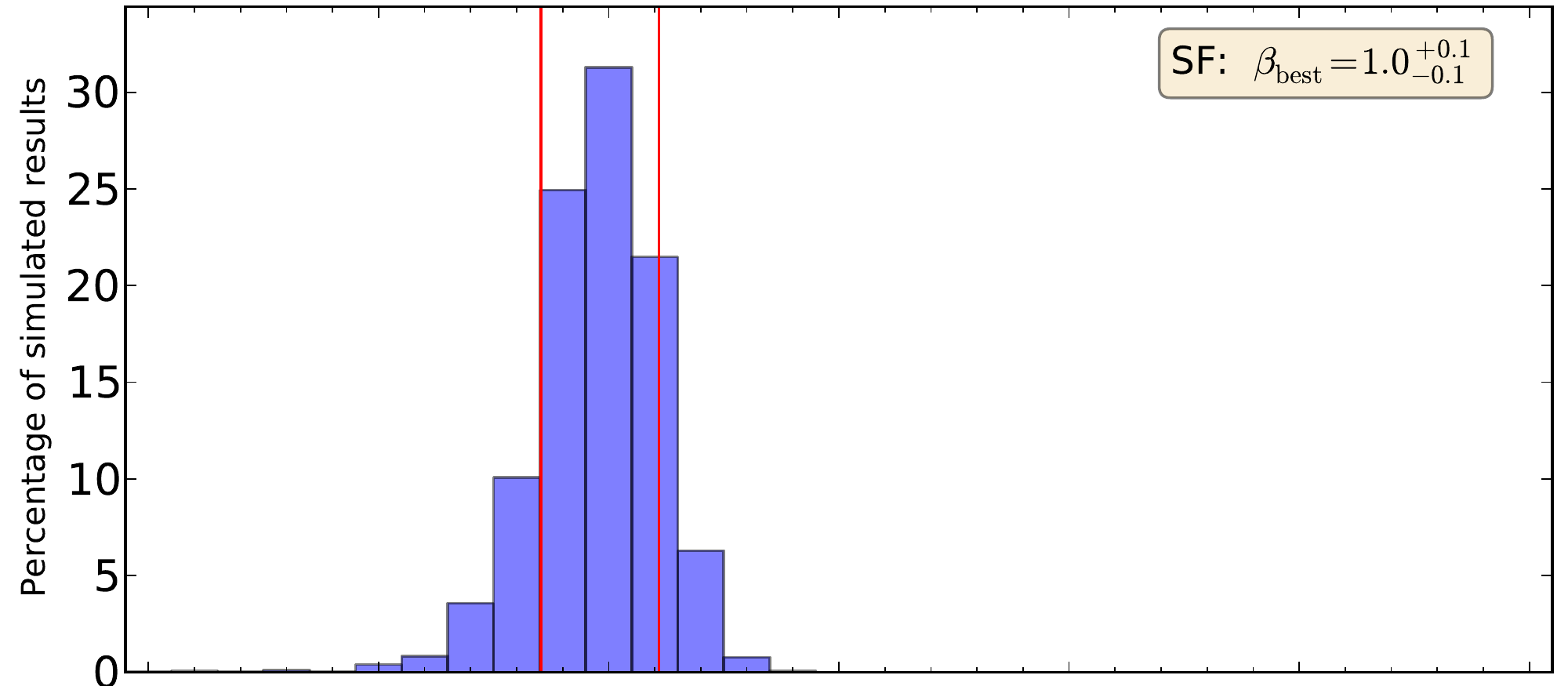}
\includegraphics[width=0.98\textwidth]{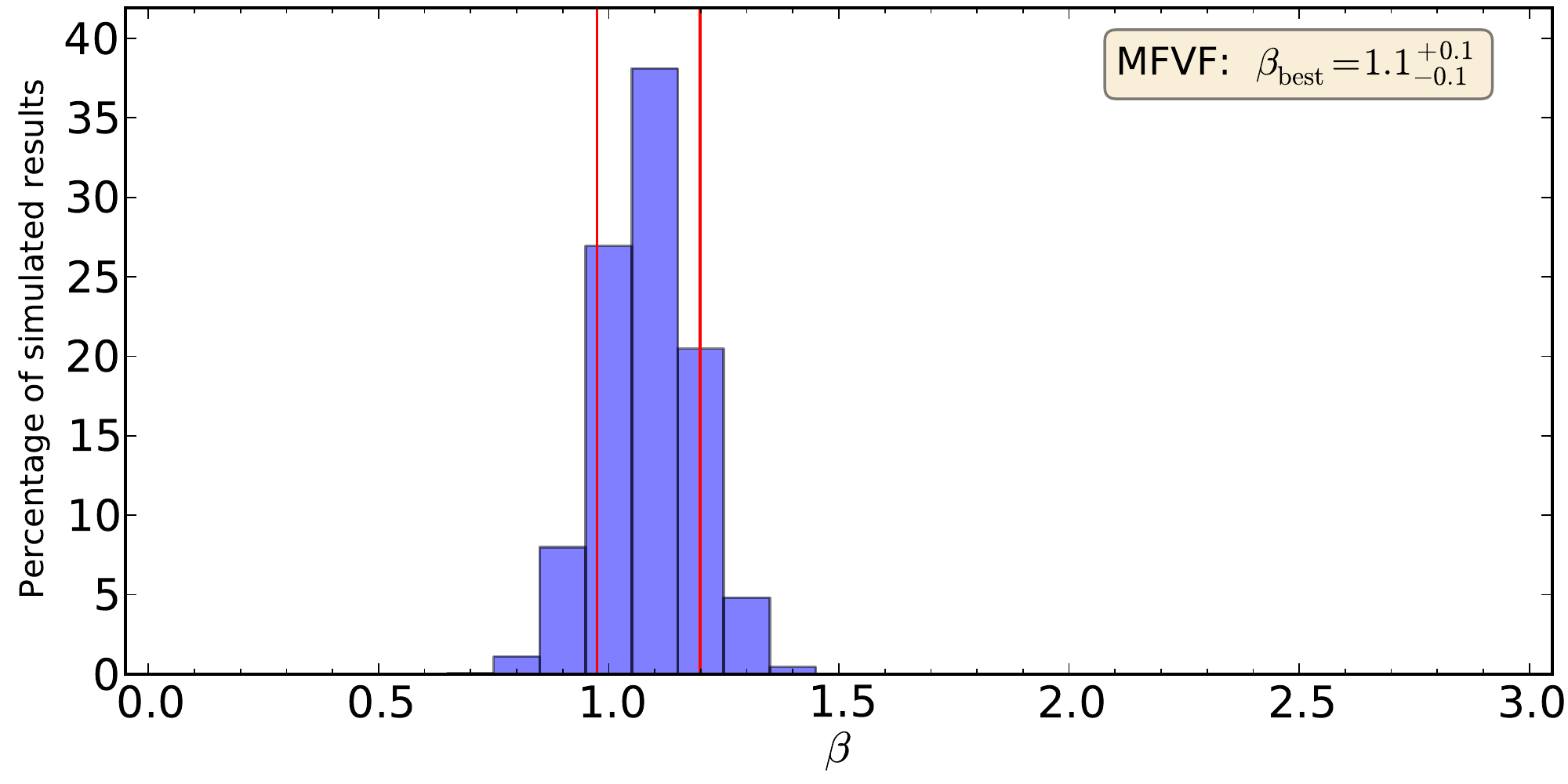}
\end{subfigure} \vline \begin{subfigure}[b]{0.49\textwidth}
\caption{\FermiLAT} \flushright
\includegraphics[width=0.98\textwidth]{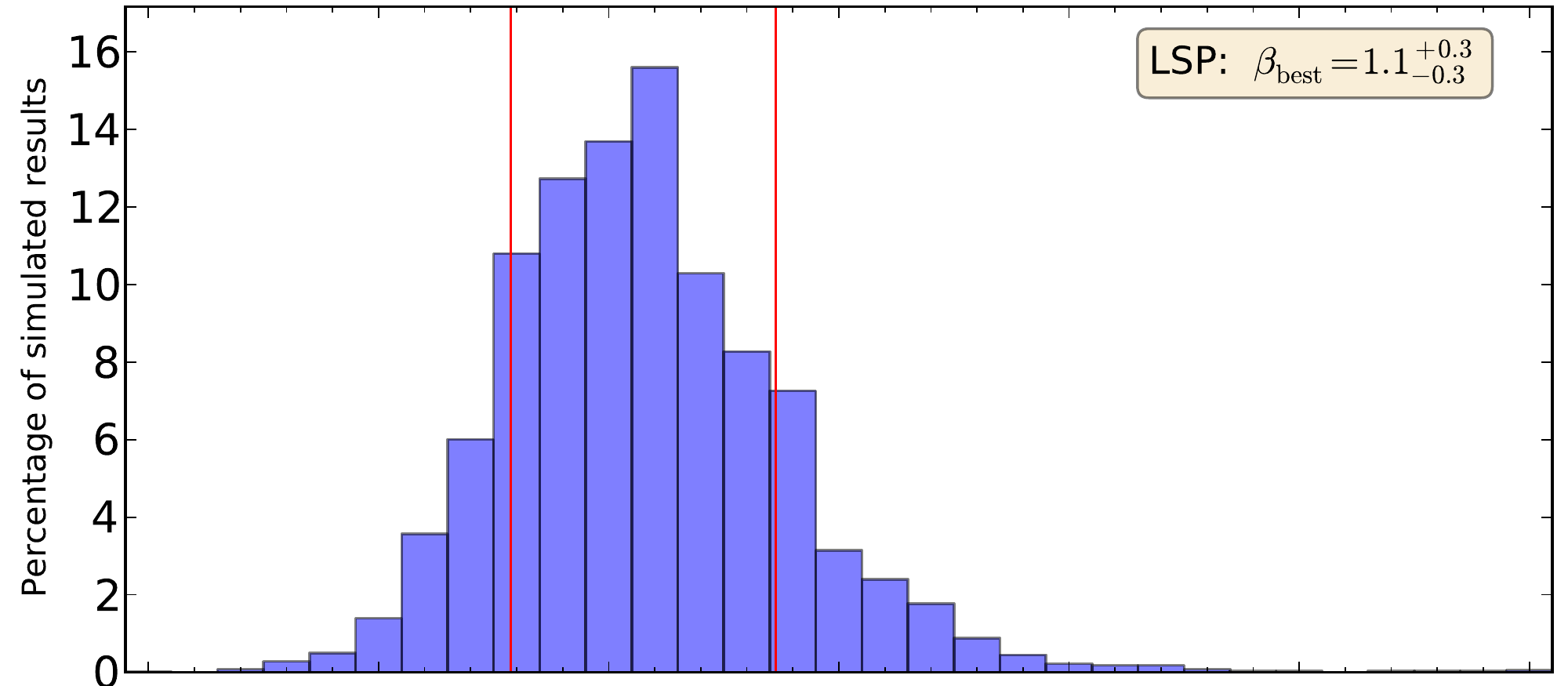}
\includegraphics[width=0.98\textwidth]{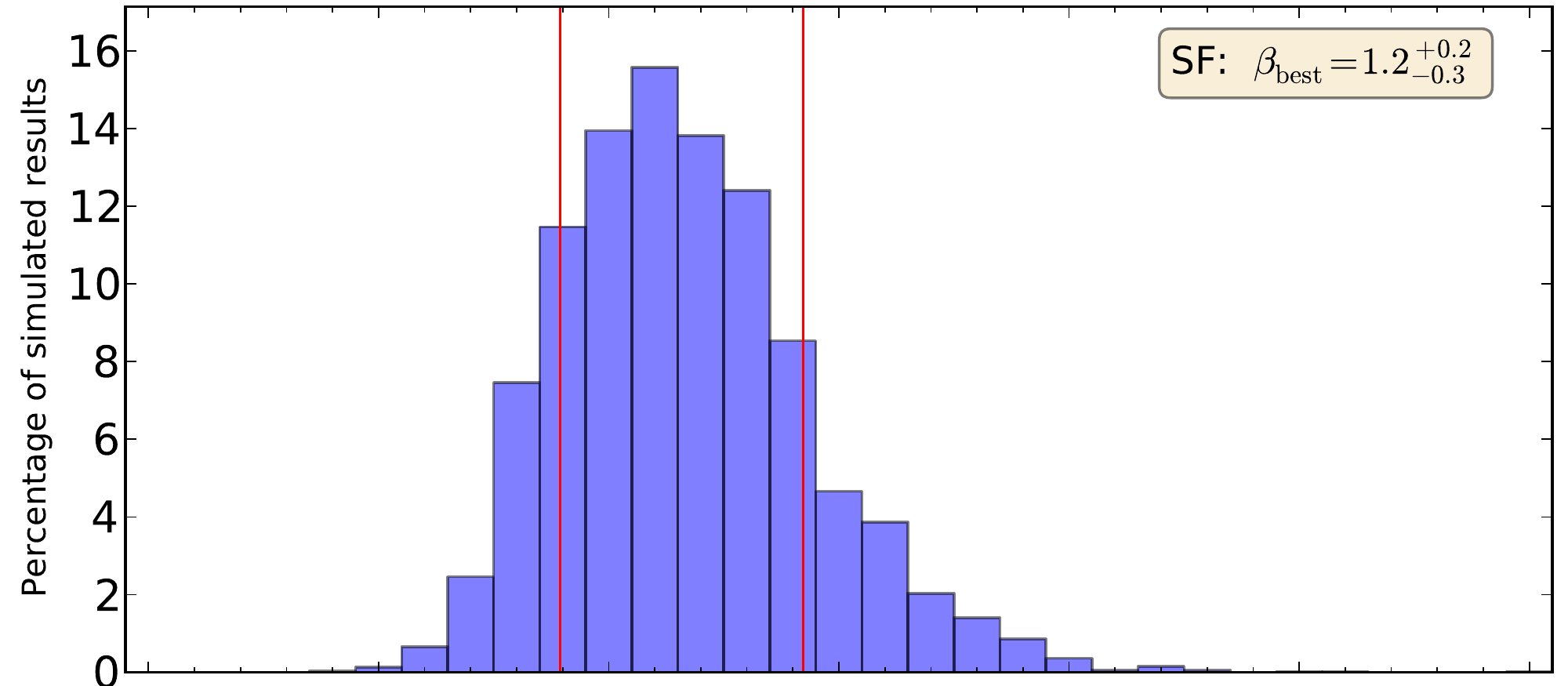}
\includegraphics[width=0.98\textwidth]{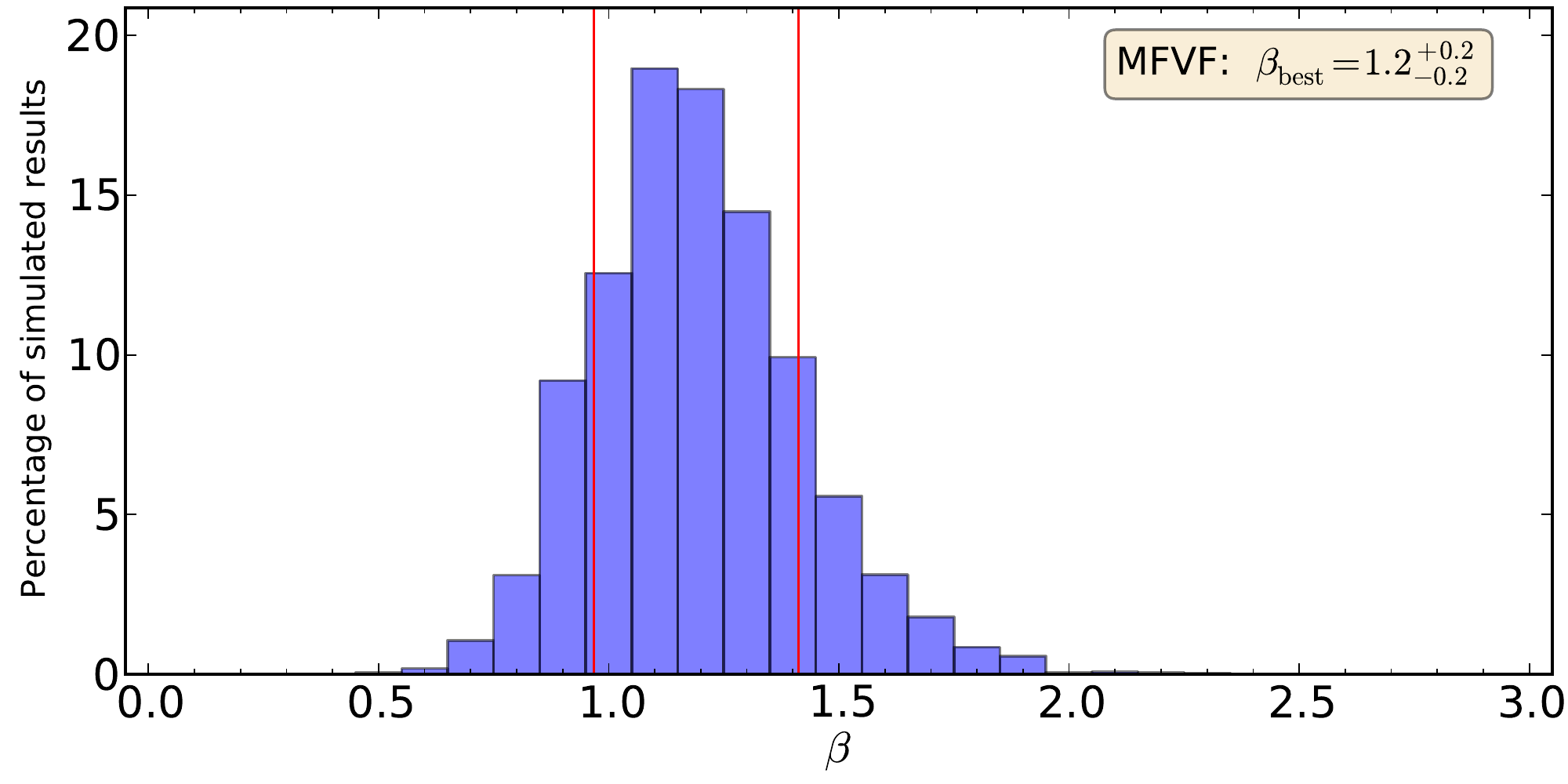}
\end{subfigure} 
\caption{Uncertainties on the LSP, SF, and MFVF best-fit parameters. \emph{Left panel (a)}: The histograms represent the distributions of estimates for $\beta$ for simulated \He light curves with the LSP (top), SF (middle), and MFVF (bottom) respectively. For the simulated light curves, the true value for $\beta$ is the best-fit value found for the \He light curve.  The vertical red bars are the $1\sigma$ uncertainties on the best fit obtained from the histograms by removing equal tails. \emph{Right panel} (b): Same plots for the \FermiLAT data.}
\label{Fig:uncertainties} 
\end{figure*}

\noindent The LSP, SF and MFVF of the measured light curves
together with the probability density functions (PDF) of the
simple power-law noise model that best fits the data are shown in
Fig.~\ref{Fig:results}. The probability is normalized to unity in
each frequency bin.  The uncertainties of $\beta$ are found with the
distributions illustrated in Fig.~\ref{Fig:uncertainties}.

\noindent For comparison, as a second hypothesis, an extended model 
is investigated: a maximum timescale is considered, above which the
variance does not increase any further. The existence of  such a
timescale is expected, as otherwise the variance of the light
curve and therefore the flux would increase to infinity with
time. This is represented by a break in the PSD to a constant
level ($\beta=0$) at the corresponding frequency $f_\mathrm{min}$,
which is treated as an additional free parameter, with a sampling
$\textrm{log}_{10}\left(f_\mathrm{min}/\mathrm{d}^{-1}\right)=-4.4,-4.2, \dots,-1.0$.

\noindent The likelihood estimators with the three methods give
best-fit values or $1\sigma$ upper limits,

$$\betahessls,$$ $$\fminhessls,$$ 
$$\betahesssf,$$ $$\fminhesssf,$$ 
$$\betahessmf,$$ $$\fminhessmf.$$

\noindent The best-fit values of
$\mathrm{log}_{10}\left(f_\mathrm{min}/\mathrm{d}^{-1}\right)=-3.80$ 
for the LSP and SF is near the formal
limit of our analysis, which is constrained to $-4.4$ by the lengths
of the simulated light curves. The results together with the $1\,\sigma$ uncertainties 
to higher values ($+1.12$ and $+1.61$) are therefore treated 
as $1\,\sigma$ upper limits. The best likelihood for the MFVF is given for a 
break in the PSD.  However, the goodness-of-fit p value of $10\%$ is not improved 
compared to $11\%$ for the first 
hypothesis (assuming no break). This and the LSP and MFVF 
findings thus do not reveal any preference for the existence of 
such a break in the sampled frequency range.

\noindent An analogous analysis is performed on the \FermiLAT light
curve. The following best-fit parameters for the simple power-law model are
compatible: 
$$\fixedbetafermils,\quad \fixedbetafermisf, \quad \fixedbetafermimf,$$ 
The most constraining result, again given by the MFVF, is taken as 
the final value: 
$$\fixedbetafermimean.$$ 
The goodness-of-fit tests yield p values $>6\%$. 
Inspection of the LSP plot in Fig.~\ref{Fig:results} for the \FermiLAT 
data reveals a peak at 
$\sim(670-700)$
days, which is indicative of a possible 
HE periodicity on the noted timescale and influencing the p value for 
this analysis. It is interesting to note that a tentative HE periodicity of 
$\sim(620-660)$ days has already been reported 
\citep[][]{2014ApJ...793L...1S}.

\noindent A comparative analysis of the extended model (power law with a break) gives compatible 
results with

$$\betafermils,$$ $$\fminfermils,$$
$$\betafermisf,$$ $$\fminfermisf,$$ 
$$\betafermimf,$$ $$\fminfermimf.$$

\noindent The best-fit values for the LSP and the SF are both
$\mathrm{log}_{10}\left( f_\mathrm{min}/\mathrm{d}^{-1}\right)=-4.20$. Including the $1\,\sigma$ 
uncertainties to higher values ($+1.42$ and $+1.40$) they are treated as $1\,\sigma$ upper
limits. 

\noindent The goodness-of-fits do not improve. The results therefore
give no indication for the presence of a break frequency in the PSD 
in the sampled frequency range.  See Table~\ref{Tab:results} for a summary of 
these and related results.

\noindent {\it Correlations:} The VHE and HE light curves are furthermore analyzed
for a possible direct correlation with the discrete correlation function 
\citep[DCF;][]{1988ApJ...333..646E} using a binning of $30\,$d. The
results are compared with the DCFs of simulated \He and \FermiLAT
light curves following one of the two hypotheses: (1)~The \He and
the \FermiLAT light curves are characterized by flicker noise
($\beta=1.1$) without any correlation.  (2)~The fluxes of such
light curves obey a perfect, direct linear correlation.  

\noindent For each hypothesis $10^4$ pairs of corresponding light 
curves are simulated and their DCFs are calculated. These DCFs 
are then combined in a two-dimensional histogram which is treated 
as a PDF.  The hypotheses are tested according to the goodness-of-fit 
test described above: for the measured DCF the likelihood is calculated 
with the PDF. Also for the simulated DCFs the likelihoods are calculated 
and are used as a test statistic for calculating the p value.  Both 
hypotheses are compatible with the measured data with p values for 
(1) of $41\%$, and for (2) of $59\%$, respectively. Accordingly, these 
results neither give a clear preference nor do they allow the rejection of a direct 
correlation.

\begin{table*}
\caption{Characteristics of the power spectral densities of \Pk at different energies and timescales.}             
\label{Tab:results} 
\centering            
\begin{tabular}{l c c l c c c c}       
\hline \hline 
Energy & Instrument & Range of & Method & $\beta$ & $\textrm{log}_{10}\left( f_\textrm{min}/\mathrm{d}^{-1}\right)$ & Goodness & Ref. \\
 range &  & $\textrm{log}_{10}\left(f/\mathrm{d}^{-1}\right)$ &  &  & & of fit & \\

\hline
$>200\, \textrm{GeV}$ & \He & $\left[-4.0, -0.3\right]$ & LSP, FF &
$\fixedbetahesslsvalues$ & no break (fixed) & 56\% & \\

$>200\, \textrm{GeV}$ & \He & $\left[-3.5, 0.0\right]$ & SF, FF &
$\fixedbetahesssfvalues$ & no break (fixed) & 48\% & \\

$>200\, \textrm{GeV}$ & \He & $\left[-3.5, 0.0\right]$ & MFVF, FF &
$\fixedbetahessmfvalues$ & no break (fixed) & 11\% & \\

$>200\, \textrm{GeV}$ & \He & $\left[-4.0, -0.3\right]$ & LSP, FF &
$\betahesslsvalues$ & $\fminhesslsvalues \fminhesslslimit$ & 51\% & \\ 

$>200\, \textrm{GeV}$ & \He & $\left[-3.5, 0.0\right]$ & SF, FF &
$\betahesssfvalues$ & $\fminhesssfvalues \fminhesssflimit$ & 43\% & \\

$>200\, \textrm{GeV}$ & \He & $\left[-3.5, 0.0\right]$ & MFVF, FF &
$\betahessmfvalues$ & $\fminhessmfvalues$ & 10\% \\

$0.1 - 300\,\textrm{GeV}$ & \FermiLAT & $\left[-4.0, -1.3\right]$ & LSP, FF & $\fixedbetafermilsvalues$ & no break (fixed) & 6.3\% & \\ 
$0.1 - 300\,\textrm{GeV}$ & \FermiLAT & $\left[-3.3, -1.0\right]$ & SF, FF & $\fixedbetafermisfvalues$ & no break (fixed) & 46\% & \\ 
$0.1 - 300\,\textrm{GeV}$ & \FermiLAT & $\left[-3.3, -1.0\right]$ & MFVF, FF & $\fixedbetafermimfvalues$ & no break (fixed) & 40 \% & \\ 
$0.1 - 300\,\textrm{GeV}$ & \FermiLAT & $\left[-4.0, -1.3\right]$ & LSP, FF & $\betafermilsvalues$ & $\fminfermilsvalues \fminfermilslimit$ & 2.4\% & \\ 
$0.1 - 300\,\textrm{GeV}$ & \FermiLAT & $\left[-3.3, -1.0\right]$ & SF, FF & $\betafermisfvalues$ & $\fminfermisfvalues \fminfermisflimit$ & 33\% & \\ 
$0.1 - 300\,\textrm{GeV}$ & \FermiLAT & $\left[-3.3, -1.0\right]$ & MFVF, FF & $\betafermimfvalues$ & $ \fminfermimfvalues$ & 30\% & \\ 

\hline 

$>200\, \textrm{GeV}$ & \He $^a$& $\left[0.9, 2.6\right]$ & SF, FF & $2.06 \pm 0.21$ & $\left[0.1, 0.9\right]$  & $-$ & (1) \\
 
$0.1-300\, \textrm{GeV}$ & \FermiLAT $^b$& $\left[-2.0, -0.9\right]$ & PSD, fit & $0.577 \pm 0.332$ & $-$  & $-$ & (2) \\ 
$0.1-300\,\textrm{GeV}$ & \FermiLAT $^c$& $\left[-3.2, -0.7\right]$ & PSD, fit & $0.64^{+0.79}_{-0.50}$ & $<-1.7$  & $-$ & (3) \\
 $2.5 - 20\,$keV & RXTE $^d$ & $\left[-0.9, 0.0\right]$ & PSD, fit & $1.46\pm 0.10$ & $-$ & $-$ & (4) \\ 
$2.5 - 20\,$keV & RXTE $^d$ & $\left[0.0, 1.9\right]$ & PSD, fit & $2.23 \pm 0.10$& $-$ & $-$ & (4) \\
Optical & Geneva & $\left[-1.2, 2.0\right]$  & SF, fit & $2.4^{+0.3}_{-0.2}$ & $-$ & $-$ & (5) \\
Optical & ROTSE & $\left[-4.4, -1.3\right]$ & MFVF, FF & $1.8^{+0.1}_{-0.2}$ & $-3.0^{+0.3}_{-0.4}$ & $-$ & (6)\\
Optical & SMARTS & $\left[-2.4, -0.9\right]$ & PSD, fit & $2.2^{+0.2}_{-0.4}$ & $-$ & $-$ & (7) \\
\hline            
\end{tabular}
\tablefoot{Range of $\textrm{log}_{10}\left(f/\mathrm{d}^{-1}\right)$: The range over which the PSD is characterized. Method: The methods used for the analysis; forward folding method (FF), best fit of the slope by, e.g. a $\chi^2$-test (fit), Structure Function (SF), Lomb-Scargle Periodogram (LSP) and Multiple Fragments Variance Function (MFVF). Goodness of fit: the p-value obtained with simulated light curves. Upper part - results from this work. Lower part - results reported in the literature with superscripts referring to: $^a$ \He flaring state: a break was detected with a $95\%$ confidence in the SF. It shall be noted that a break in the SF is at a $\sim 3$ times larger frequency than in the intrinsic PSD. $^b$ Based on aperture photometric \FermiLAT light curves provided by the Fermi Science Support Center at \url{http://fermi.gsfc.nasa.gov/ssc/data/access/lat/msl\_lc/}. $^c$ Model results assuming a superposition of Ornstein-Uhlenbeck processes (OU) and using 4 years of \FermiLAT data. A slight preference for a single OU with different slope is reported. $^d$ Based on non-simultaneous data.}
\tablebib{(1)~\citet{2010A&A...520A..83H}, (2)~\citet{2013ApJ...773..177N}, (3)~\citet{2014ApJ...786..143S}, (4)~\citet{2001ApJ...560..659K}, (5)~\citet{1997A&A...327..539P}, (6)~\citet{2011A&A...531A.123K}, (7)~\citet{2012ApJ...749..191C}.}
\end{table*}



\section{Discussion and conclusions}
\label{Sec:Discussion}
For the first time the temporal variability of the VHE emission of \Pk 
in the quiescent state has been analyzed on timescales from days up 
to more than nine years.
The variability of the long-term quiescent VHE light
curve as measured with \He provides evidence for a lognormal
behavior and is compatible with power-law noise process on timescales 
$>1$~d.  The VHE PSD on these timescales is consistent with a 
power law ($\propto f^{-\beta}$) with an index of $\fixedbetahessmean$ 
(flicker noise). On the other hand, the PSD for the \He data during 
the flaring period in 2006 is consistent with a power law of slope 
$\beta=2.06 \pm 0.21$ (red noise) on timescales $<3\,\textrm{h}$ 
with indications for a possible break in the SF between $3$ and $20\,$h
\citep{2010A&A...520A..83H}.  In the context of accretion-powered
sources, X-ray variability with similar characteristics (power-law noise 
with $\beta \sim 1-2$, and lognormal behavior) has often been related 
to random fluctuations of the disk parameter $\alpha$ on local viscous timescales
 \citep[e.g.,][]{1997MNRAS.292..679L,2004MNRAS.348..111K}. 
The current VHE findings can in principle be interpreted in two different 
ways:

\noindent (1) The PSD slope of the VHE variability is stationary.
It follows a power law with a transition (break) from $\beta
\sim2$ to $\sim1$  somewhere between $0.1$ and $\sim1$~d. 
This can be compared to the X-ray PSDs of Seyfert AGN and radio
galaxies. The PSD of Seyfert AGN exhibit a power-law behavior
 $\beta \simeq 1$ at low frequency, steepening
to $\beta \simgt 2$ on timescales shorter than some break time
$T_{br}$ \citep[e.g.,][]{2006Natur.444..730M}. Two radio galaxies,
3C 111 and 3C 120 show a similar behavior in X-ray
with a power-law slope of $\sim 2$ for 3C 111 \citep{2011ApJ...734...43C} and a steepening of 
the PSD at high frequency for 3C 120 \citep{2009ApJ...704.1689C}. Interestingly, for \Pk a 
break time $T_{br} \sim1$~d has been suggested earlier \citep[][cf also
Tab.~1,]{2001ApJ...560..659K} based on (nonsimultaneous) X-ray
data \citep[cf. also][for caveats]{2010MNRAS.404..931E}.  In the
case of Seyfert AGN a simple quantitative relationship between
$T_{\rm br}$, the observed (bolometric) luminosity $L_B$ and the black
hole mass $M_{\rm BH}$ has been found \citep[][]{2006Natur.444..730M}. 
Although \Pk is not a radio-quiet object, a similar relation could apply 
if its characteristic timing properties are caused by an external
process \citep[e.g., originate in the accretion flow, see for
instance;][]{2010LNP...794..203M}. In terms of the accretion rate
$\dot{m}_E$ (expressed in units of the Eddington rate), the
relevant scaling relation for a standard disk configuration
becomes $(T_{br}/1~\rm{d})  \simeq 0.7~(M_{\rm BH}/10^8
M_{\odot})^{1.12}/ \dot{m}_E^{0.98}$. If this would apply to \Pk,
then relatively high accretion rates ($\simgt 0.1$ Eddington rate
for $M_{\rm BH} \geq 10^8 M_{\odot}$) would be implied even for
the quiescent VHE state. This may suggest that in this context
the break time is more likely related to a change in accretion flow 
conditions such as a transition from an advection-dominated to
a standard disk configuration.

\noindent (2) Alternatively, the power-law indices of the PSD
could be different during the quiescent and flaring
states, as they are possibly related to different physical processes
and/or spatial locations. The extreme characteristics of the flaring
state (including apparent lognormality and a red-noise behavior
down to minutes) seem to require rather exceptional conditions to
account for it
\citep[e.g.,][]{2010A&A...520A..23R,2012A&A...548A.123B,2012MNRAS.420..604N}.
This could support a variability origin differing from the origin in
the quiescent state. 
\noindent Further limits on the possible cross-over timescale
will be important to distinguish between these two scenarios.

\noindent The HE light curve, as measured with \FermiLAT, is found
to be compatible with a power-law noise of slope
$\fixedbetafermimean$ on timescales larger than ten days. This 
value differs slightly from the earlier reported best-fit $\beta=1.7\pm
0.3$ for the average PSD of the six brightest \FermiLAT BL Lacs 
(including \Pk) based on the first 11 months of data \citep{2010ApJ...722..520A}. 
It is compatible with more recent indications for a flatter slope
($\beta \sim 1$) for high-frequency-peaked BL Lac (HBL) objects
(S. Larsson, private communication). The HE
slope is also compatible with the VHE results for the quiescent
data set, suggesting that the PSDs on these timescales are
shaped by similar processes. 
\noindent Owing to instrumental noise and observational gaps a 
possible direct correlation between the VHE and the HE light 
curves could neither be excluded nor firmly established.

\noindent The HE and VHE PSDs show a scale invariance on timescales 
from weeks up to at least the $1\sigma$-lower limits $\simgt 
600\,$d and $\simgt 200\,$d, respectively. 
A maximum timescale of $10^3$ days has been reported in the
optical range (see Tab.~1). 
If the maximum timescale was related to the radial infall 
time in the accretion disk, then a possible outer radius of
$r_d \simeq(\alpha \sqrt{GM_{\rm BH}} f_{\rm min}^{-1})^{2/3} \simgt
4\times10^{16} \ (\alpha/0.3)^{2/3}(t_{\rm max}/1000\
\mathrm{d})^{2/3}$ $(M_{\rm BH}/10^{8} M_{\odot})^{1/3}\,$cm
might be inferred for an advection-dominated system
\citep{1997MNRAS.292..679L}. However a significant detection of
such a maximum timescale in the $\gamma$-ray range will probably
require longer light curves that exceed this time-scale by an order of
magnitude.

\noindent For both data sets of \Pk, the flux distributions during the 
quiescent state are compatible with lognormal distributions, where the
result is more significant for the VHE data than for the HE data. A hint
of lognormal behavior was found in \cite{2010A&A...520A..83H} where the 
distribution of the fluxes of the quiescent state of 2005-2007 were following a 
lognormal distribution.
This suggests that multiplicative, i.e., self-boosting processes dominate 
the variability. It is interesting to note that in the context of galactic 
X-ray binaries, where lognormal flux variability has first been established, 
such a behavior is thought to be linked to the underlying accretion 
process \citep{2001MNRAS.323L..26U}. In the AGN context, evidence 
for lognormality on different timescales has in the meantime also been 
found in several sources, for example, in the X-ray band for the BL~Lac object BL~Lacertae 
\citep{2009A&A...503..797G}, in the TeV band for the BL~Lac object Markarian~501 
\citep{2010A&A...524A..48T, 2015arXiv150904893C}, and in the X-ray band for 
the Seyfert~1 galaxy IRAS~13~244-3809 \citep{2004ApJ...612L..21G}. 


\noindent Further observations with \HeII and the planned
Cherenkov Telescope Array will allow us
to improve the characterization of the PSD at high frequencies
due to their better sensitivities \citep{2013APh....43..215S}.  The larger energy range will
make it possible to close the gap to the \Fermi band, helping to
improve our understanding of the degree of convergence (e.g.,
possible correlations and similar processes) between the HE and
the VHE domain.  A clear characterization of the $\gamma$-ray
variability of different sources similar to \Pk will also be
important to improve our understanding of the physical processes
separating different source classes.


\begin{acknowledgements}
The support of the Namibian authorities and of the University of Namibia in facilitating the construction and operation of H.E.S.S. is gratefully acknowledged, as is the support by the German Ministry for Education and Research (BMBF), the Max Planck Society, the German Research Foundation (DFG), the French Ministry for Research, the CNRS-IN2P3 and the Astroparticle Interdisciplinary Programme of the CNRS, the U.K. Science and Technology Facilities Council (STFC), the IPNP of the Charles University, the Czech Science Foundation, the Polish Ministry of Science and Higher Education, the South African Department of Science and Technology and National Research Foundation, the University of Namibia, the Innsbruck University, the Austrian Science Fund (FWF), and the Austrian Federal Ministry for Science, Research and Economy, and by the University of Adelaide and the Australian Research Council. We appreciate the excellent work of the technical support staff in Berlin, Durham, Hamburg, Heidelberg, Palaiseau, Paris, Saclay, and in Namibia in the construction and operation of the equipment. This work benefited from services provided by the H.E.S.S. Virtual Organisation, supported by the national resource providers of the EGI Federation.
\end{acknowledgements}


\bibliographystyle{aa}
\bibliography{referenzen}

\end{document}